\documentclass[twocolumn,showpacs,superscriptaddress,groupedaddress]{revtex4-1}

\usepackage{graphicx}  
\usepackage{dcolumn}   
\usepackage{bm}        
\usepackage{amssymb}   
\usepackage{physics}
\usepackage[caption=false]{subfig}
\usepackage{tabularx,ragged2e,booktabs,caption}
\usepackage{cellspace}
\usepackage{color}
\newcolumntype{C}[1]{>{\Centering}m{#1}}

\begin{document}

\widetext

\title{Gravimetry through non-linear optomechanics}

\author{Sofia Qvarfort}
\affiliation{Department of Physics and Astronomy, University College London, Gower Street, WC1E 6BT London, United Kingdom}
\author{Alessio Serafini}
\affiliation{Department of Physics and Astronomy, University College London, Gower Street, WC1E 6BT London, United Kingdom}
\author{Peter F. Barker}
\affiliation{Department of Physics and Astronomy, University College London, Gower Street, WC1E 6BT London, United Kingdom}
\author{Sougato Bose}
\affiliation{Department of Physics and Astronomy, University College London, Gower Street, WC1E 6BT London, United Kingdom}

\date{\today}

\begin{abstract}
We propose a new method for measurements of gravitational acceleration using a quantum optomechanical system. As a proof-of-concept, we investigate the fundamental sensitivity for a cavity optomechanical system for gravitational accelerometry with a light-matter interaction of the canonical `trilinear' radiation pressure form. The phase of the optical output of the cavity encodes the gravitational acceleration $g$ and is the only component which needs to be measured to perform the gravimetry.  We analytically show that homodyne detection is the optimal readout in our scheme, based on the cyclical decoupling of light and matter, and predict a fundamental sensitivity of $\Delta g = 10^{-15}$ ms$^{-2}$ for currently achievable optomechanical systems which could, in principle, surpass the best atomic interferometers even for low optical intensities. Our scheme is strikingly robust to the initial thermal state of the mechanical oscillator as the accumulated gravitational phase only depends on relative position separation between components of the entangled optomechanical state arising during the evolution.
\end{abstract}

\maketitle

Inertial sensors are an important and integral component of our current technological society. The practise of measuring the gravitational acceleration $g$ -- also known as gravimetry -- has lead to important advances in both fundamental science and industry. For example, local variations of $g$ have been mapped with the GRACE satellite to construct global tidal models \cite{ray2006tide}, and more recently, the Juno spacecraft measured the gravity harmonics of Jupiter \cite{iess2018measurement}. Furthermore, precise measurements of $g$ can test for small deviations from Newtonian gravity on extremely small scales, which may provide indications of a deeper theory of quantum gravity \cite{biswas2012towards}. In industry, precision accelerometry is extensively used in inertial navigation and for conducting geological surveys. 

While classical systems have long been utilised to perform accurate measurements of $g$, quantum systems offer several useful advantages, including reduced noise levels, a compact realisation and most importantly an increased measurement sensitivity achieved through the power of coherence and interferometry. Over the past decade, a variety of quantum systems have been explored to this aim, in both theory and practice. The largest research effort to date has focused on atom interferometry \cite{peters2001high,mcguirk2002sensitive,bidel2013compact,hu2013demonstration}, for which the highest achieved sensitivity currently stands at $\Delta g = 4.3\times 10^{-9}$ ms$^{-2}$ \cite{hu2013demonstration}. A similar investigation has been carried out for both on-chip and fountain Bose-Einstein condensate (BEC) interferometry with best sensitivity $\Delta g  = 7.8 \times 10^{-10}$ ms$^{-2}$ \cite{abend2016atom}. Finally, a proposal for using magnetically levitated spheres which predicts sensitivities of $2.2 \times 10^{-9}$  ms$^{-2}$Hz$^{-1/2}$ has been put forward in \cite{johnsson2016macroscopic}. For comparison, the current commercial standard is set by the LaCoste FG5-X gravimeter which can achieve a measurement sensitivity of $1.5\times 10^{-9}$ ms$^{-2}$Hz$^{-1/2}$ \cite{LaCoste2016}. More generally, the broader topic of using quantum systems to probe relativistic phenomena is currently being pursued with great interest (see for example \cite{dimopoulos2008general,bruschi2014quantum, howl2016gravity, seveso2016can, seveso2017quantum,  asenbaum2017phase, tan2017relativistic, joshi2017space}). 

\begin{figure}[h]
   \subfloat[]{
            \begin{minipage}{0.5\linewidth} \label{fig:MirrorCavity}
                \includegraphics[width=0.99\linewidth, height = 0.2\textheight, keepaspectratio=true]{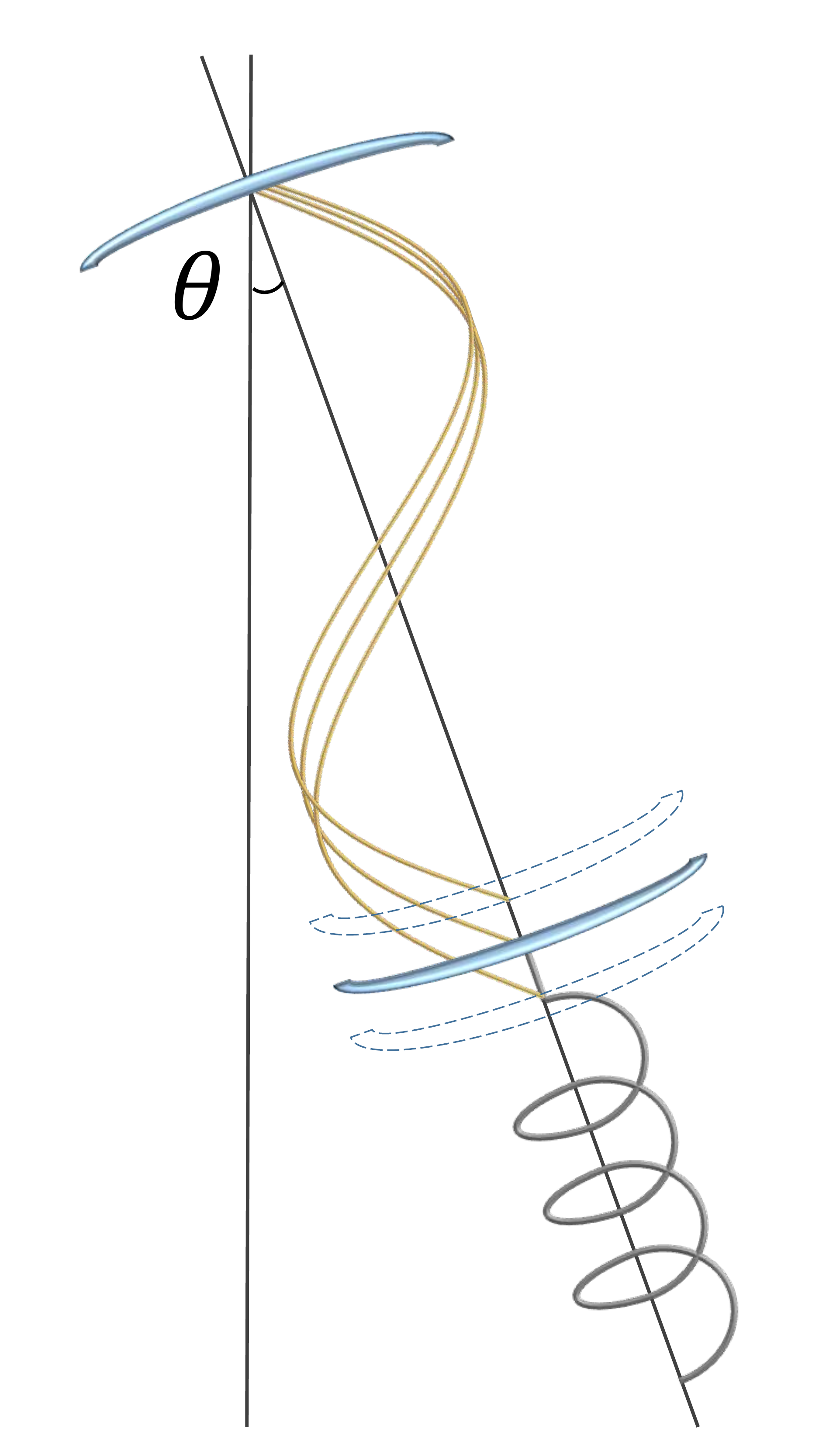}
            \end{minipage}} 
    \subfloat[]{
            \begin{minipage}{0.5\linewidth} \label{fig:Quadratures}
                \includegraphics[width=0.98\linewidth, height = 0.2\textheight, keepaspectratio=true]{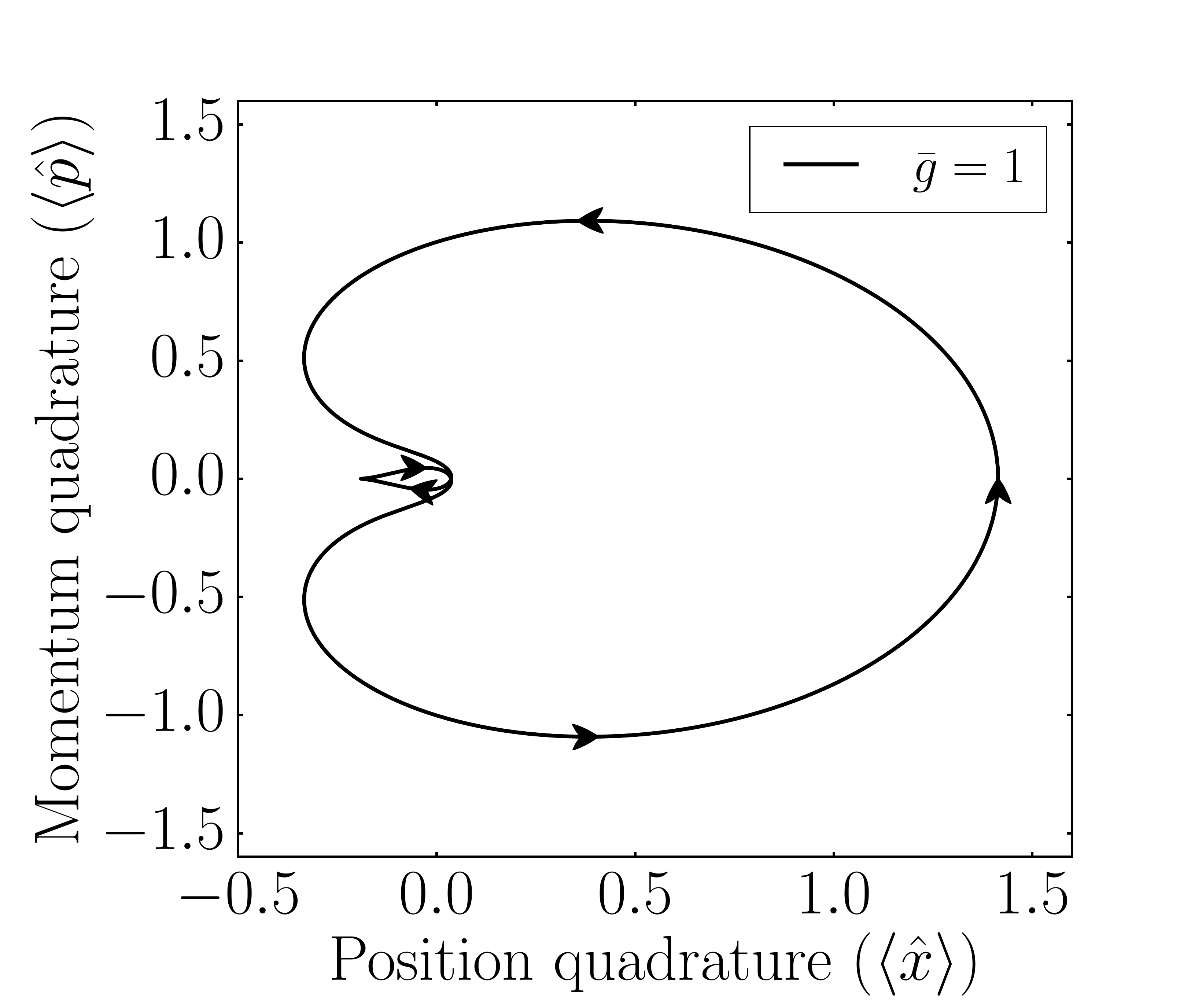}
                \includegraphics[width=0.98\linewidth, height = 0.2\textheight, keepaspectratio=true]{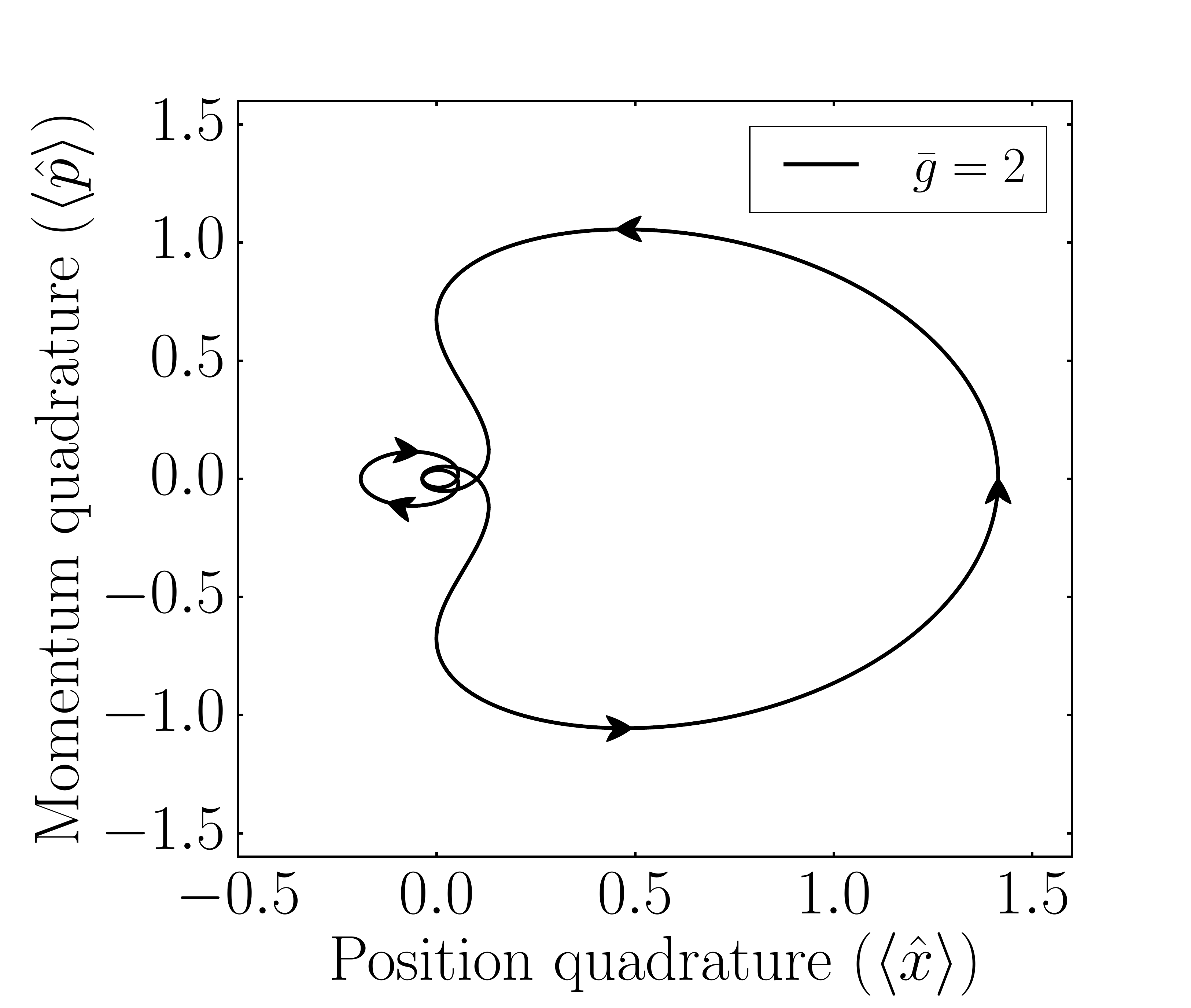}
            \end{minipage}}  \hfill
\caption{Figure showing (a) an inclined cavity with a moving mechanical mirror which is displaced by gravity $g$ and (b) the quadratures of the cavity state for parameters $\bar{k} = 1, \bar{g} = 1$ and $\bar{k} = 1, \bar{g} = 2$,  both using $\alpha =\beta = 1$.}  
\label{fig:System}
\end{figure}

A key advantage to quantum systems are their interferometric properties. In order to further improve the measurement sensitivity, one can ask how these interferometric properties can be further enhanced. One way to achieve this is to place a quantum system in the form of a mechanical oscillator in an optical cavity, a research area known as quantum optomechanics  \cite{aspelmeyer2014cavity}. The addition of the cavity allows for a strong coherent coupling between light and oscillator which, as we shall see, cancels out any initial thermal noise and fundamentally improves the measurement sensitivity of the device. 

Within classical optomechanics, the idea of gravimetry and accelerometry by optically detecting the mechanical oscillator has been experimentally realised by Cervantes in\textit{et al}. \cite{Cervantes2014}. Other avenues, such as the detection of high frequency gravitational waves through the driving of resonant mechanical elements was proposed also in \cite{Arvanitaki2013}. In the related field of electromechanics, Schr\"{o}dinger cat states and a Kerr nonlinearity have recently been found to be useful for the same applications \cite{jacobs2016quantum}. However, the ensuing fundamental limits on the measurement sensitivity of gravimetry in the `quantum' regime of optomechanics using its trilinear radiation pressure interaction is yet to be investigated. Here we undertake this task and obtain some striking results: (i) It is possible, in principle, to surpass the sensitivity $\Delta g $ that has been obtained in atom interferometers and other implementations to date, (ii)  due to the periodic decoupling of light and mechanics, the mechanical element does not require initial cooling to the ground state to improve the fundamental sensitivity of the gravimeter and, finally, (iii) the best possible sensitivity is achieved by a simple homodyne measurement of the cavity field, while only a low photon number in the cavity is required. That is, no measurement on the mechanical oscillator is required. Unlike the case of atomic interferometers, in optomechanics the interaction of light and matter is continuous, and we will see that our Hamiltonian cyclically entangles and disentangles the light and mechanics, leading to their decoupling. It follows that the experimental challenge will be to maintain the quantum coherence of the field and mechanics over the duration of each run of the experiment, which we set as one oscillation period of the mechanical element.  This requirement, on which the plausibility of the scheme hinges, will be discussed in some detail. 

The paper is organised as follows. We first propose an optomechanical Hamiltonian with a gravitational potential energy associated with the vertical position of the oscillator. Then, starting with the light and the oscillator as coherent states, we derive an analytical expression for the state evolution and show that a measurement of $g$ only requires probing the cavity state. We will also show that the sensitivity after a single oscillation period is impervious to the initial thermal state of the oscillator. We then present our main results, which involve deriving analytical expressions for the quantum and classical Fisher information for measurements of $g$. In addition, we prove that at the time where light and mechanics disentangle, the optimal measurement is given by a homodyne detection scheme on the cavity state. Using state-of-the-art parameters, we consider two distinct optomechanical systems and calculate the resulting ideal $\Delta g$ for each of them.  We will also attempt to estimate the precision that can be obtained in the lab by evolving the system numerically in a very narrow parameter range and then using our results as guidance for what to expect of realistic systems. Finally, we will discuss the experimental feasibility of this scheme and provide a comprehensive comparison between our results and other currently available gravimetry proposals. 

\textit{The system}. 
Let us begin by considering a general optomechanical system consisting of a mechanical oscillator coupled to a light field in the cavity. The non-gravitational Hamiltonian that describes the dynamics of an optomechanical system is given by \cite{mancini1997ponderomotive,bose1997preparation}:
\begin{equation} \label{eq:Hamiltonian}
\hat{H} = \hbar \omega_C a^\dagger a + \hbar \omega_m b^\dagger b - \hbar k a^\dagger a (b^\dagger + b), 
\end{equation}
where $a, a^\dagger$ are the annihilation and creation operators for the cavity field with frequency $\omega_C$, $b, b^\dagger $ are the annihilation and creation operators for the mechanical oscillator with frequency $\omega_m$, and $k$ (usually denoted with $g$ in the literature, but which we shall here reserve for gravity) is a coupling constant that determines the interaction strength between the photon number $a^\dagger a $ and the position $\hat{x}_O \propto (b^\dagger + b)$ of the oscillator. While we will keep the subsequent discussion general, let us here provide three examples of common optomechanical systems and their respective coupling constants. For a a Fabry-Perot cavity with a mechanical oscillator mirror, $k$ takes the form \cite{mancini1997ponderomotive,bose1997preparation}
\begin{eqnarray} 
k_{\mathrm{FP}} = \frac{\omega_C}{L} \sqrt{\frac{\hbar }{2m \omega_m}},  \label{eq:MirrorCavity}
\end{eqnarray}
where $L$ is the length of the cavity and $m$ is the mass of the mirror. A levitated nano- or micro-crystal (e.g. a diamond or silicon bead), on the other hand, has a $k$ given by \cite{chan2011laser, chang2010cavity}
\begin{eqnarray}
k_{\mathrm{Lev}} =  \frac{P}{4V_C\epsilon_0}  \sqrt{\frac{\hbar }{2m \omega_m}} k_C \omega_C, \label{eq:NanoDiamond}
\end{eqnarray} 
where $\epsilon_0$  is the permittivity of free space,  $V_C$ is the cavity mode volume,  and $k_C$ is the wave-vector of the laser, given by $2\pi/\lambda$, where $\lambda$ is the laser wavelength. $P= 3V\epsilon_0(\epsilon - 1)/(\epsilon + 2)$ is the polarizability of the levitated object of volume $V$ and $\epsilon $ is the relative electric permitivity.  Alternately, we can also consider a BEC trapped in a cavity. Here, the collective motion of the ensemble acts as the massive oscillator. For this system, the  coupling constant is given by \cite{brennecke2008cavity, munstermann1999dynamics}
\begin{equation}
k_{\mathrm{BEC}} = \frac{\sqrt{N} g_0^2 k_1}{\Delta_{ca}} \sqrt{\frac{\hbar}{2 M \omega_m}},
\end{equation}
where $N$ is the number of atoms in the ensemble, $g_0$ is the single-atom cavity QED coupling rate, $M = N m $ is the collective mass of all the trapped atoms with individual mass $m$, $k_l $ is the wave-vector of the laser and  $\Delta_{\mathrm{ca}} = \omega_p - \omega_C$ with pumping frequency $\omega_p$. We will return to these expressions when computing the fundamental sensitivity limits for each system in the latter part of the paper.
 
In order to introduce a coupling to a gravitational potential in the Hamiltonian, we add a term of the form $ m g \hat{x}_O  \cos{\theta}$. Here, $m$ is the mass of the mechanical oscillator, $g$ is the gravitational acceleration, $\hat{x}_O = \sqrt{\hbar/2m\omega_m}(b^\dagger + b)$ is the position operator acting on the mechanical oscillator, and $\theta$ is an angle from the vertical axis that we include in order to describe inclined systems \cite{Scala2013a}.  For example, a Fabry-Perot mirror-cavity system can be inclined from the vertical as seen in Fig. \ref{fig:MirrorCavity}.  Note that while the mass $m$ appears as a coupling in the Hamiltonian, we will later see that measurements of $g$ are mass-independent, which is what we expect from the equivalence of inertial and gravitational mass.  
The Hamiltonian of the system thus becomes
\begin{align} \label{eq:GravHamiltonian}
\hat{H}_g = &\hbar \omega_C a^\dagger a + \hbar \omega_m b^\dagger b - \hbar k a^\dagger a ( b^\dagger + b) \nonumber \\
&+ \cos{\theta}g  \sqrt{\frac{\hbar m }{2\omega_m}}(b^\dagger +b) . 
\end{align}

\textit{System dynamics}. 
In order to simplify the time evolution operator $U(t)$  corresponding to the above Hamiltonian, we rescale $H_g$ by dividing all terms by the oscillator frequency $\omega_m$. As a result, the time parameter $t$ now represents the labframe time  multiplied by $\omega_m$, such that the oscillator has undergone a full oscillation cycle at $t = 2\pi$. The operator $U(t) $ can then be written in the following simplified form (see Appendix A in \cite{bose1997preparation} for details of the derivation in the absence of gravity):
\begin{eqnarray} \label{eq:Time}
U(t) &=\exp{- i r a^\dagger a t} \exp{ i ( \bar{k} a^\dagger a -  \bar{g})^2 (t - \sin{t}) } \nonumber \\
& \times\exp{(\bar{k} a^\dagger a - \bar{g})( \eta b^\dagger - \eta^* b )] \mathrm{exp}[- i b^\dagger b t}, 
\end{eqnarray}
where $r = \omega_C/\omega_m$, $\eta = 1 - e^{-it}$, $\bar{k} = k/\omega_m$, and $\bar{g} = \cos{\theta}g \sqrt{ m/(2\hbar  \omega_m^3)}$.  As a rule, we will denote any dimensionless quantity with a bar. For time-dependent variables, such as dissipation rates, this means they have been rescaled with respect to $\omega_m$. 

We now assume that the cavity field mode and the mechanical oscillator are initially in coherent states $\ket{\alpha}_C$ and $\ket{\beta}_O$ respectively. For laser light injected into the cavity, this is the natural assumption. The oscillator, on the other hand, will in reality be initialised as a thermal state, which corresponds to a random coherent state $\ket{\beta}_O$ according to a thermal distribution. However, by starting out with a coherent state we will later argue that the gravimetric phase accumulated by the light does not depend on $\ket{\beta}_O$ so that our procedure works equally well for an arbitrary thermal state. A formal proof of this statement can be found in the Appendix. The initial state  at $t = 0$ is then given by $
\ket{\Psi(0)} = \ket{\alpha}_C \otimes \ket{\beta}_O$, and under $U(t)$ it gives us the following state
\begin{align} \label{eq:state}
\ket{\Psi(t) } = &e^{- |\alpha|^2/2} \sum_{n = 0}^\infty \biggl[ \frac{\alpha^n}{\sqrt{n!}} e^{i (\bar{k}^2 n^2 - 2 \bar{k} \bar{g} n ) \tau} \nonumber \\
&\times e^{( \bar{k}n - \bar{g} )\left( \eta^* \beta - \eta \beta^*\right)/2}\ket{n}_C \otimes \ket{\phi_n(t)}_O\biggr],
\end{align}
where $\tau = t - \sin{t}$, and $\ket{\phi_n(t)}_O$ are coherent states of the oscillator given by $\ket{\phi_n(t)}_O = \ket{e^{-it} \beta + (\bar{k} n - \bar{g} )(1 - e^{-it})}$.  In the derivation of this state, we have adopted a rotating frame for the cavity field, thus ignoring the free evolution  induced by the term $\exp{- i ra^\dagger a}$. 

The state in Eq. (\ref{eq:state}) show us that light and mechanics will entangle and disentangle  periodically, with maximum entanglement occurring at $t = \pi$. At $t = 2\pi$, the oscillator state $\ket{\phi_n}_O$ returns to $\ket{\beta}_O$  regardless of the values of $\bar{k}, \bar{g}$ and $\beta$, and therefore by extension a thermal state also returns to its initial state because it will undergo the same compact evolution. This means that the initial oscillator state does not impact the fundamental sensitivity of this scheme. As already mentioned, a formal proof of this can be found in the Appendix. Most importantly however, at $t = 2\pi $ the cavity state is completely decoupled from the oscillator, meaning that all information about $g$ is transferred to the phase of the cavity state.  As a result, any measurement scheme needs only consider the cavity state after one oscillation period, meaning that direct or indirect access to the oscillator state is not required. This is exactly why the scheme allows one to measure the cavity state exclusively in order to infer $g$, which greatly simplifies an experimental implementation. This convenient property arises from the interferometric properties of the oscillator; its quantum nature allows it to acts as an interferometer to ensure that any \textit{initial} thermal noise is removed from the cavity field, and thereby our scheme does not require cooling of the oscillator to a pure ground state. In other words, our results are valid for both coherent and thermal states. Note however that decoherence ensuing from damping to the oscillator motion \textit{during} the state evolution will adversely affect the final measurement sensitivity and cause the oscillator state to grow increasingly mixed. We will not consider this kind of decoherence in this work, and instead assume that the mechanical element remains coherent over one oscillation period. 

We can visualise some of the dynamics of the state in Eq. (\ref{eq:state}) by computing the expectation values of the field quadratures $\hat{X}_C = ( a + a^\dagger )/\sqrt{2}$ and $\hat{P}_C= i (a^\dagger - a )/\sqrt{2}$ \cite{serafini2017}. We focus on the cavity state, which we obtain by tracing out the oscillator. It is given by 
\begin{align} \label{eq:TracedCavity}
\rho_C(t) = &e^{-|\alpha|^2} \sum_{n, n'}^\infty \biggl[  \frac{\alpha^n(\alpha^*)^{n'}}{\sqrt{n!n'!}} e^{i \left( \bar{k}^2(n^2 - n^{'2}) - 2\bar{k}\bar{g}(n-n') \right)\tau}  \nonumber \\
&\times e^{\left(\bar{k}(n-n') - \bar{g}\right)\left( \eta\beta - \eta^* \beta \right)/2}\nonumber \\
&\times e^{- |\phi_{n}|^2/2 -  |\phi_{n'}|^2 / 2 + \phi^*_{n'} \phi_n} \ket{n} \bra{n'} \biggr].
\end{align}
For decoherence-free evolution, the trajectories traced out by the system in phase space can be seen for different values of $\bar{g}$ in Fig. \ref{fig:Quadratures}. We find that the system performs increasingly complex trajectories for larger values of $\bar{g}$.

\textit{Gravimetry}. 
We now come to our main results which concern the use of optomechanical systems as gravimeters. The question we wish to answer is: what is the best fundamental sensitivity $\Delta g$ with which an optomechanical system can measure the gravitational acceleration $g$? $\Delta g$ here denotes the standard deviation of a gravimetric measurement. We can directly predict $\Delta g$ from the system's dynamics by calculating the Fisher information $I_F$. The Fisher information provides a natural lower bound on the variance $\mathrm{Var}(g)$ of an unknown parameter, in our case $g$. This relationship is captured by the Cram\'{e}r-Rao inequality \cite{cramer1946contribution, rao1992information, wolfowitz1947efficiency} 
\begin{equation} \label{eq:CramerRao}
\mathrm{Var}( g )\geq \frac{1}{I_F}.
\end{equation}
Thus if we maximise $I_F$, we minimise the measurement spread of $g$. 

\textit{Quantum Fisher Information}. 
The Fisher information comes in two forms: the measurement-specific classical Fisher information (CFI) and the quantum Fisher information (QFI). The QFI, which we denote $H_Q$, is computed by optimising over all possible POVMs and their resulting CFI \cite{helstrom1976quantum, holevo2011probabilistic}. Thus  $H_Q$ represents the ultimate bound on obtainable information from a system, but it does not reveal which specific measurement is required to achieve it. For a general mixed quantum state $\rho(t)$ the QFI is given by $ H_Q(t) = \trace{[\rho(t) \mathcal{L}^2]}$, 
where $\mathcal{L}$ is the symmetric logarithmic derivative. In general, it is very difficult to obtain $\mathcal{L}$ analytically, although there are methods for finding a noisy bound on the Cram\'{e}r-Rao inequality \cite{alipour2014quantum}. A similar method for many-body systems was proposed in \cite{beau2017nonlinear}, and numerical methods were shown to be effective for a class of specific systems \cite{liu2016quantum}. We shall not be using these methods here, as we shall instead investigate specific measurements for the noisy scenario to better approximate an experimental setting. This will later allow us to prove the optimality of the homodyne measurement. 

Let us start by deriving a fundamental bound to the sensitivity.  We specialise to the simpler case where the state $\rho(t)$ is pure. Setting $\rho(t) = \ketbra{\Psi(t)}$ the quantum Fisher information becomes
\begin{eqnarray} \label{eq:HQ}
H_Q(t) = 4 \left[ \braket{\partial_g\Psi(t)} - |\langle \Psi(t)|\partial_g \Psi(t) \rangle|^2 \right], 
\end{eqnarray}
where we have used the notation $\partial_g = \partial/\partial g$. 

At first glance, the QFI of the global system might not seem very relevant as the mechanical part of the optomechanical system cannot easily be measured directly. However, we recall that the coherent state $\ket{\phi_n(t)}_O$ returns to $\ket{\beta}_O$ at $t = 2\pi$, so that all information about $g$ is transferred to the phase of the pure, decoupled cavity state. Since the decoupling time does not depend on $\beta$, this is also the case for a thermal state that may be written as a statistical mixture of coherent states (see the Appendix for a proof of this statement). Calculating the QFI for this state will therefore provide an experimentally accessible notion of the fundamental sensitivity of the device. We find  the following expression for $H_Q$ at $t = 2\pi$: 
\begin{equation} \label{eq:QuantumFisher}
H_Q(2\pi) = \frac{32 \pi^2 \bar{k}^2 m |\alpha|^2 \cos^2{\theta} }{\hbar \omega^3_m}.
\end{equation}
Note that the mass term $m$ is canceled by the appearance of $m$ in the coupling constant $\bar{k}$, so that the final accelerometry measurement is independent of $m$. Note also the strong dependence on $\bar{k}$ and $\omega_m$, and that the expression scales linearly with the number of photons $|\alpha|^2$. 

\begin{figure}
\subfloat[ \label{fig:QFI}]{%
  \includegraphics[width=0.49\linewidth]{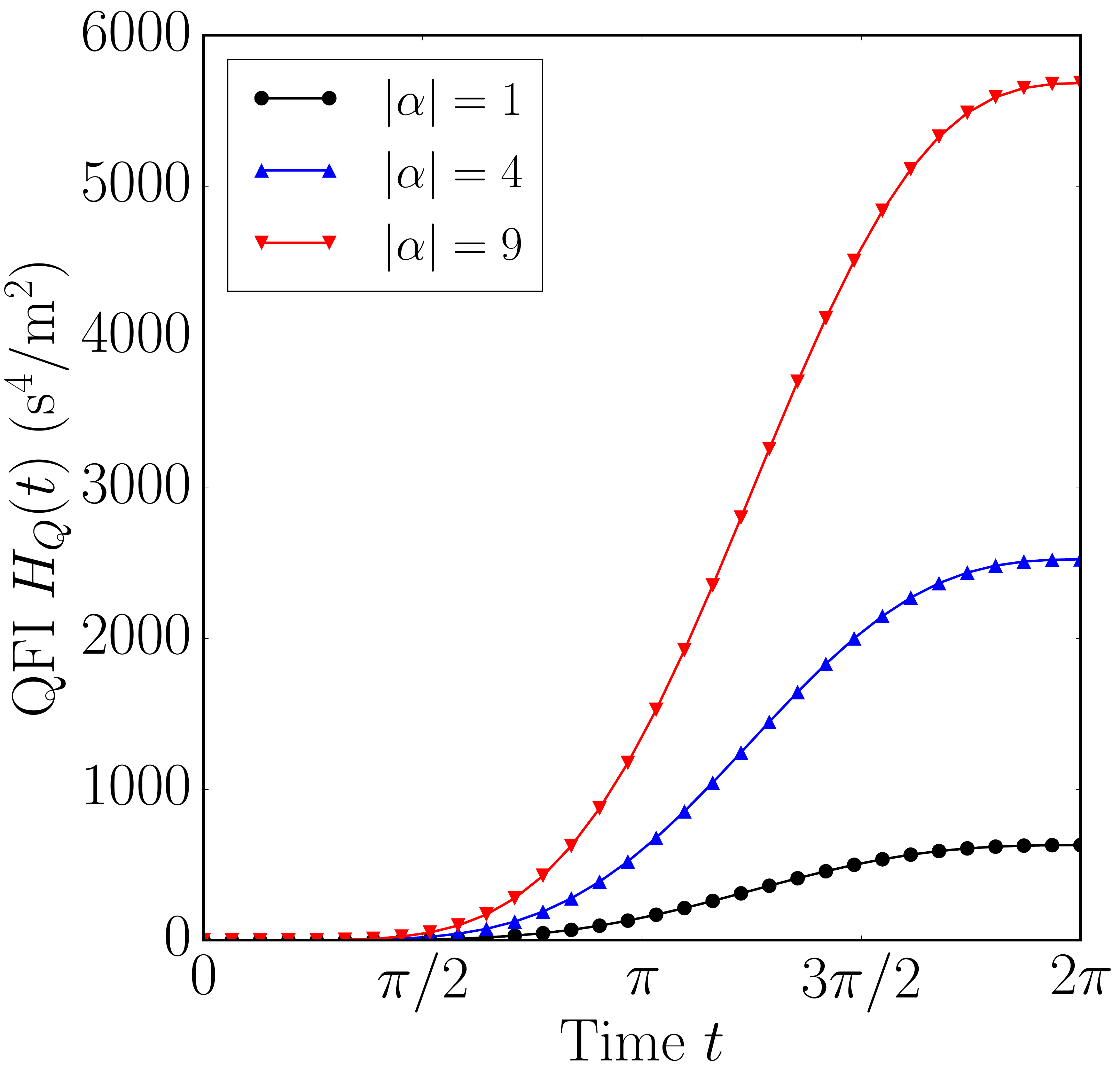}%
  } \hfill
    \subfloat[ \label{fig:CFIrescaled}]{%
  \includegraphics[width=0.49\linewidth]{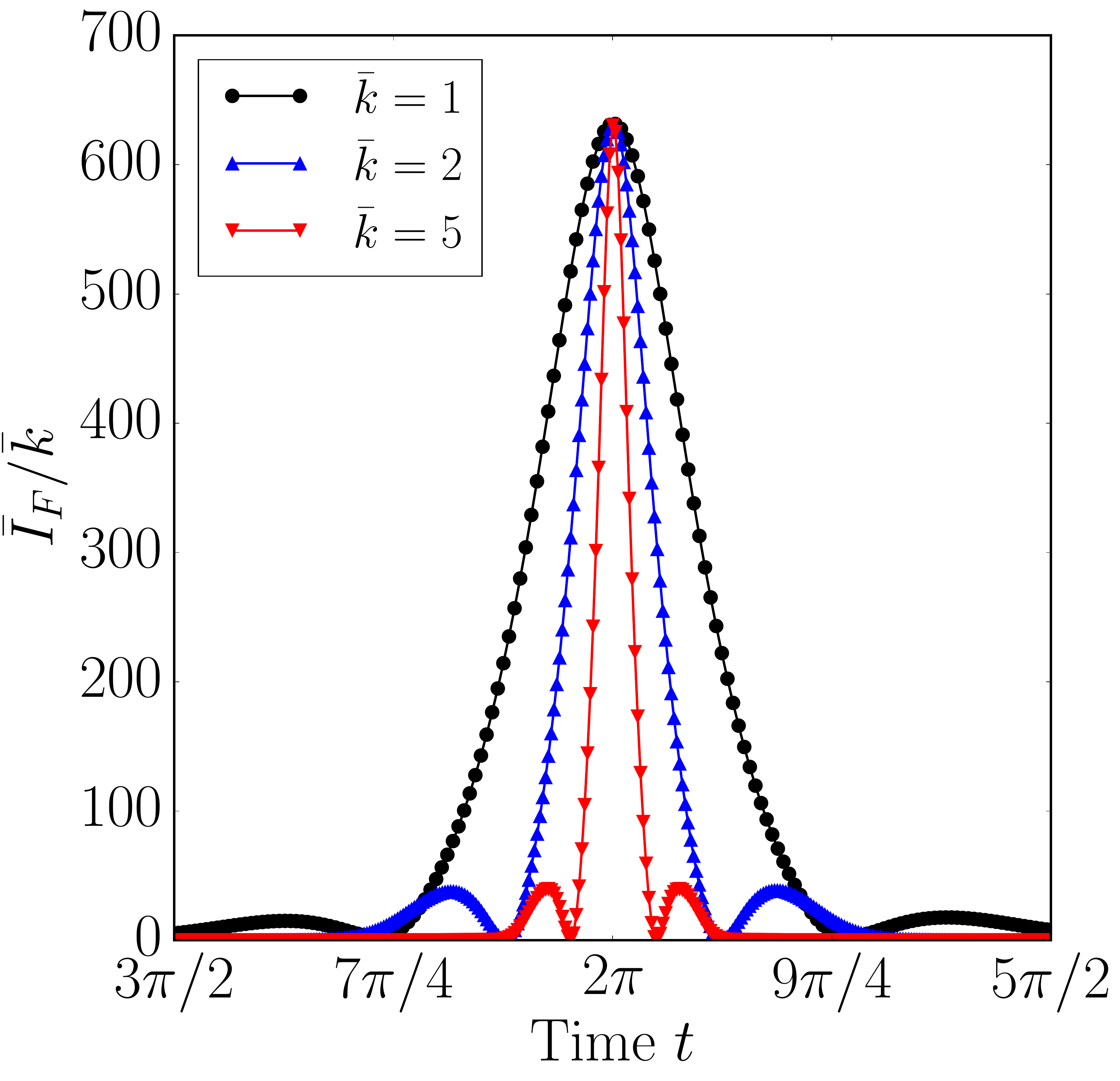}%
  } \hfill
    \subfloat[ \label{fig:CFIMomentum}]{%
  \includegraphics[width=0.49\linewidth]{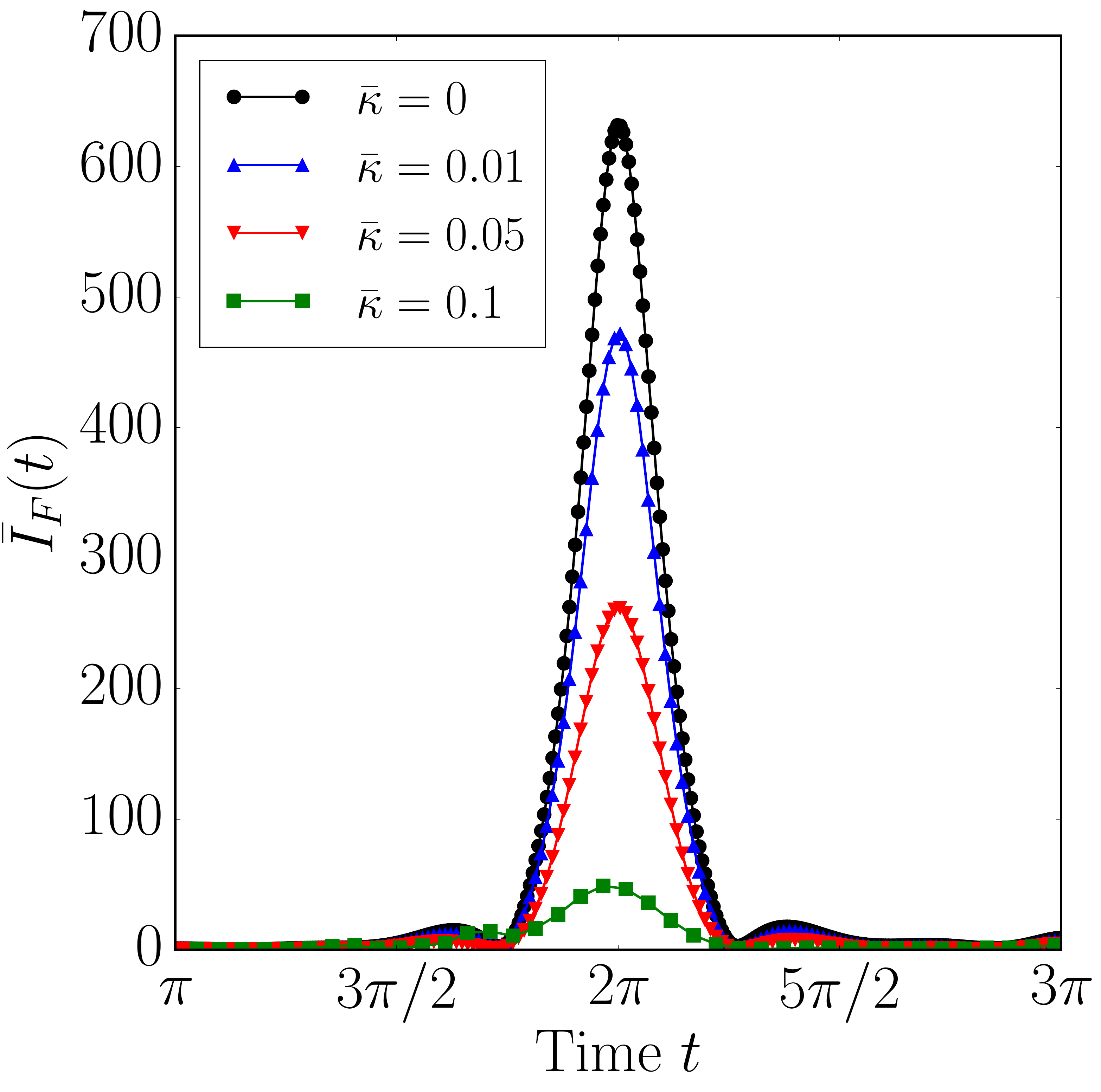}%
  } \hfill
\subfloat[ \label{fig:Leaky}]{%
  \includegraphics[width=0.49\linewidth]{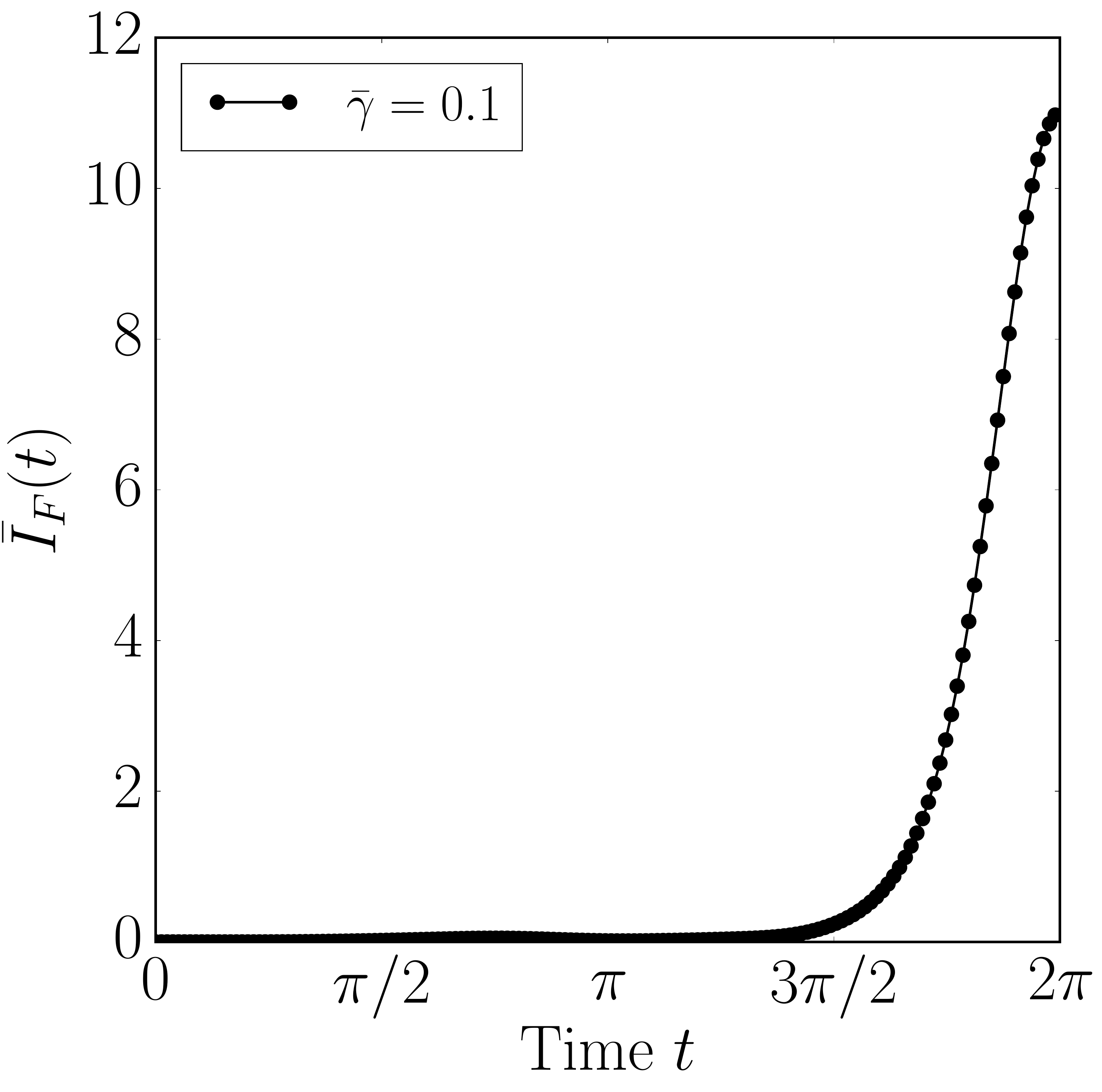}%
} 
\caption{Plots of the quantum and classical Fisher information for measurements of $g$.  (a) shows the QFI $H_Q$ for $1$, $4$ and $9$ photons with $\bar{k} = \bar{g} = 1$. (b) shows how the CFI for a momentum measurement narrows with increasing $\bar{k}$ for $\bar{g} = \alpha = \beta =1$, (c) shows the CFI for a momentum measurement with various decoherence rates $\bar{\kappa}$ for $\bar{k} = \bar{g} = \alpha = \beta = 1$, and (d) shows the $I_F$ for a momentum measurement of photons that leak from the cavity with coupling $\bar{\gamma} = 0.1$ and $\bar{k} = \bar{g} = \alpha = \beta = 1$. }
\label{fig:Fisher}
\end{figure}

To find $H_Q(t)$ for the global state at any time $t$ we resort to numerical calculations. We consider the case $|\alpha|^2 = 1$ and set $\bar{k} = \bar{g} = 1$ to allow for future comparisons with subsequent numerical evaluation of the CFI which will be restricted to the same narrow parameter range. The resulting $H_Q(t)$ as a function of $t$ can be found in Fig. \ref{fig:QFI}. We note that $H_Q(t)$ reaches its maximum value at $t = 2\pi$, which means that Eq. (\ref{eq:QuantumFisher}) returns the largest possible value during one oscillation period for any choice of system. 

\textit{Classical Fisher Information}. 
Let us now consider a specific measurement of $g$. The classical Fisher information (CFI) $I_F$ determines the minimum standard deviation of a parameter estimator once we have chosen a single specific measurement with POVM elements $\{\Pi_x\}$. The CFI is given by the expression
\begin{eqnarray}
I_F(t) = \int \mathrm{d} x\frac{1}{p(x, g) } \left( \frac{\partial p(x, g)}{\partial g} \right)^2, \label{eq:CFI}
\end{eqnarray}
where $p(x, g) = \trace{[\Pi_x \rho(g)]}$ is a probability distribution that we obtain by measuring $\Pi_x$ on the state $\rho(g)$. 

We now consider a general homodyne measurement on the traced-out cavity state $\rho_C$, corresponding to the Hermitian operator 
$\hat{x}_\lambda = \left( a \exp{- i \lambda} + a^\dagger \exp{i \lambda } \right)/\sqrt{2}$, where $\lambda$ denotes a label that rotates between the field quadratures \cite{barnett2002methods}. 
Any two operators that differ by $\lambda = \pi/2$ form a conjugate pair which satisfies the position-momentum commutator relation. 
In the following, we shall refer to the choices $\lambda=0$ and $\lambda=\pi/2$ as a `position' or `momentum' measurement respectively.  In order to calculate $I_F(t)$ we must find the probability distribution $p(x_\lambda, g) = \trace{[\ketbra{x_\lambda}  \rho_C(g)]}$, where $\ketbra{x_\lambda}$ are the eigenstate of $\hat{x}_\lambda$. While the position eigenstates themselves are not proper vectors, we can make use of a standard result from the quantum harmonic oscillator:  $\langle n | x_\lambda \rangle = \pi^{-1/4} 2^{-n/2} (n!)^{-1/2} \exp{-x^2_\lambda /2 }  H(x_\lambda) \exp{in\lambda}$  \cite{barnett2002methods}, to write
\begin{align}
p(x_\lambda, g) =  e^{- |\alpha|^2} \sum_{n,n'} &\biggl[ \frac{\alpha^n(\alpha^*)^{n'} }{\sqrt{n!n'!} } e^{i \left( \bar{k}^2(n^2 - (n')^2) - 2 \bar{k}\bar{g}(n-n')\right) \tau} \nonumber \\
&\times  \frac{e^{- x_\lambda^2} }{\pi^{1/2}}  \frac{H_n(x_\lambda ) H_{n'}(x_\lambda ) e^{- i \lambda  (n-n')}}{2^{(n+n')/2} \sqrt{n!n'!}}  \nonumber \\
&\times e^{\left(\bar{k}(n-n') - \bar{g}\right)\left( \eta\beta - \eta^* \beta \right)/2}\nonumber \\
&\times e^{- |\phi_n|^2/2 - |\phi_{n'}|^2/2 + \phi_{n'}^* \phi_n } \biggr],
\end{align}
where $H_n(x)$ are the Hermite polynomials of order $n$. These probabilities in turn gives rise to a CFI of the form
\begin{align} \label{eq:CFIHomodyne}
I_F(  t) =&\cos^2{\theta} \frac{m}{2\hbar \omega_m^3}\left( - 4\bar{k}^2 \tau^2 \right) e^{-|\alpha|^2} \nonumber \\
&\times \int \mathrm{d}x_\lambda \frac{\left[ \sum_{n,n'} (n-n') c_{n,n'} d_{n,n'}(x_\lambda) f_{n,l}\right]^2}{\sum_{n,n'} c_{n,n'} d_{n,n'}(x_\lambda) f_{n,n'} },
\end{align}
where
\begin{subequations}
\begin{align}
c_{n,n'} &= \frac{(\alpha^*)^{n'} \alpha^n }{\sqrt{n!n'!}} e^{i \left[\bar{k}^2 (n^2 - n^{'2} ) - 2 \bar{k}\bar{g} (n-n') \right] \tau}, \\
d_{n,n'}(x_\lambda) &= \frac{e^{- x_\lambda^2} }{\pi^{1/2}}  \frac{H_n(x_\lambda ) H_{n'}(x_\lambda ) e^{- i \lambda  (n-n')}}{2^{(n+n')/2} \sqrt{n!n'!}}, \\
f_{n,n'} &=   e^{\left(\bar{k}(n-n') - \bar{g}\right)\left( \eta^* \beta - \eta \beta^* \right)/2 }\nonumber \\
&\quad \times e^{- |\phi_{n'}|^2/2 - |\phi_n|^2/2 + \phi_{n'}^* \phi_n }.\label{eq:coefficients}
\end{align} 
\end{subequations}
Let us attempt to analyse the expression for $I_F(t)$. We immediately note that any terms in the sum with $n = n'$ do not contribute to the Fisher information. The remaining behaviour of $I_F$ can be inferred from the second exponential in $f_{n,n'}$, namely $\mathrm{exp}\{ -|\phi_{n'}|^2/2 - |\phi_n|^2/2 + \phi_{n'}^*\phi_n  \}$ as this will dominate the entire expression for large $\bar{k}$. If we simplify the expression in the exponential, we find that it is equal to
\begin{equation}
\mathrm{exp}\left\{- \bar{k}^2 (n-n')^2 (1 - \cos{t}) + \frac{\bar{k}(n-n')}{2}  \left[ \beta \eta - \beta^* \eta^* \right]  \right\}
\end{equation}
For $n \neq n'$ and large $\bar{k}$, the first term will dominate, and the exponential will be small for any $t$ that is not a multiple of $2\pi$. In other words, the Fisher information for a homodyne measurement becomes significant only when light and mechanics are completely decoupled. Fig. \ref{fig:CFIrescaled} shows how the CFI for a momentum measurement (with $\lambda = \pi/2$) for  $\bar{g} = \alpha = 1$ and $\bar{k} = 1,2,5$ becomes increasingly narrow as $\bar{k}$ grows larger. For clarity, we have rescaled  $I_F$ with $\bar{k}$ in the plot. Note that for small $\bar{k}$ we still find large $I_F$  at times $t\neq 2\pi$. See the Appendix for additional plots detailing this behaviour.  

We saw earlier that the QFI scales with $\bar{k}^2$, which mean the scheme favours systems with a large single-photon coupling. We shall soon show that the CFI coincides with the QFI at $t = 2\pi$, but in the meantime we must explore what the narrowing of the CFI at $t = 2\pi$ entails for our measurement scheme. Ultimately it will require the homodyne measurement to be performed within an increasingly narrow time-window. We can estimate the timescale in question by finding the full-width-half-maximum (FWHM) of the peak. To do so, we consider only the dominant first term $- k^2 (n-n')^2 (1 - \cos{t})$ for small perturbations in $t$ around $t = 2\pi$, thus $\cos{(2\pi + t')} \approx 1 - t^{'2}/2$. That brings the first term into the form $- k^2 (n-n')^2 (t')^2/2$, which is now a Gaussian distribution. For a Gaussian function with $\mathrm{exp} \left[ - (x-x_0)^2/(2\sigma^2) \right]$, the full-width-half-maximum (FWHM) is given by $2 \sqrt{2 \ln{2}} \sigma$. In our case, we find $\sigma^2 = [2\bar{k}^2 (n-n')^2]^{-1}$. We already noted that terms with $(n-n')$ will not contribute to the CFI, and any term with $|n-n'|\gg 1$ will just cause the peak to narrow further. Thus we only consider the terms with $|n-n'| = 1$, leaving us with $\sigma = [2\bar{k}]^{-1}$, and so we conclude that any measurement must be performed roughly on a timescale of $(\omega_m \bar{k})^{-1} = k^{-1}$. 

Let us see if we can simplify the expression for $I_F$ even further and whether it bears any semblance to the QFI. At $t = 2\pi$, $\phi_n(2\pi ) = \beta$ and $\eta = 0$. Then setting $\bar{k}$ and $\bar{g}$ to integer values causes $I_F(2\pi)$ to lose all dependence of $\bar{g}$. The coefficients reduce to $c_{n,n'} = (\alpha^*)^{n'} \alpha^n /\sqrt{n!n'!}$  and $f_{n,n'} = 1$. We now consider the generating function for the Hermite polynomials $e^{2xt - t^2} = \sum_{n = 0}^\infty  t^n H_n/n!$.  Taking the derivative of both sides results in $(2x - 2t)e^{2xt - t^2} = \sum_{n = n'}^\infty t^{n-n'} H_n /(n-n')!$, which we can use to show that   Eq. (\ref{eq:CFIHomodyne}) reduces to the compact expression
\begin{equation} \label{eq:HomodyneFisher}
I_F(2\pi) =\frac{8 \pi^2 \bar{k}^2 m}{\hbar \omega_m^3}  (i e^{-i \lambda } \alpha - i e^{i \lambda} \alpha^*)^2 . 
\end{equation}	
This expression coincides precisely with the QFI in Eq. (\ref{eq:QuantumFisher}) for complementary choices of $\lambda$ and $\alpha$. To better see why, we rewrite the term in the brackets as $\left[ ( e^{- i \lambda } - e^{i \lambda} )i \Re{\alpha} - (e^{- i \lambda } + e^{i \lambda} ) \Im{\alpha} \right]^2$.  We now note that when $\lambda = 0 $, only $\Im{\alpha}$ contributes to the CFI, whereas at $\lambda = \pi/2$, only $\Re{\alpha}$ contributes. For both of these specific choices of $\lambda$,  the CFI coincides precisely with the QFI in Eq. (\ref{eq:QuantumFisher}) because the term in the brackets reduces to $4\Re{\alpha}^2$ or $4 \Im{\alpha}^2$, respectively. We conclude that the homodyne measurement saturates the quantum Fisher information limit up to a phase dependence of $\alpha$, which can always be accounted for by changing the quadrature of the homodyne measurement.  Note however that at other times than $t = 2\pi$. the homodyne measurement will be zero for all choices of $\lambda$ and $\alpha$, and so it only saturates the QFI at the moment of decoupling. 

Finally, the absence of $\bar{g}$ from $I_F(2\pi)$  is not a problem for sensing $g$ -- it just means that the sensitivity at times $t=2\pi$ is independent of the actual value of $g$. Numerical analysis suggests that larger values for $\bar{g}$ causes the CFI to oscillate increasingly quickly before reaching its maximum value (see the supplemental material the Appendix). The optimality of the homodyne detection for sensing within our scheme is greatly advantageous as it is a routine measurement which is easy to accomplish. It has in fact also been shown to be an optimal measurement \cite{latmiral2016probing} in other contexts.

\textit{Decoherence}. The calculation above is valid for pure states, but in practice every measurement will suffer various forms of decoherence. We will here investigate the effects of decoherence on the classical Fisher information for a narrow parameter range, as realistic parameters are very difficult to simulate numerically. We shall later use these results as indications of the behaviour of realistic systems.  

There is a large variety of decoherence effects for optomechanical systems, such as decoherence due to photons leaking from the cavity, or phonons gradually being lost from the mechanical element. The latter manifests as a gradual damping of the oscillator motion, which moves the state towards a mixture in the coherent state basis \cite{zurek1981pointer, paz1993reduction, anglin1996decoherence, zurek1993coherent}. This problem has previously been treated analytically, which is possible because the decoherence operators commute with the Hamiltonian. Thus we refer to these works and will not treat the mechanical decoherence here. Instead, we make the assumption that the phonon decoherence is negligible over one oscillation period of the oscillator.

The effect of photons leaking from a cavity on a state $\rho(t)$ can be modeled using a Lindblad master equation of the form
\begin{equation} 
\frac{\partial \rho(t)}{\partial t} = - \frac{i}{\hbar} [H, \rho(t)] +  L \rho(t) L^\dagger - \frac{1}{2} \left\{ \rho(t),  L^\dagger L \right\},  \label{eq:Lindblad}
\end{equation}
where $\{\cdot, \cdot\}$ denotes the anti-commutator, $L = \sqrt{\bar{\kappa}} a$ are Lindblad operators,  and $\bar{\kappa}$ is the decoherence rate with respect to the rescaled time $t$, as opposed to the lab-time decoherence rate $\kappa$. This equation cannot easily be solved analytically since the operator $a$ does not commute with the Hamiltonian $H_g$ in Eq. (\ref{eq:GravHamiltonian}). Some solutions have been found for specific cases, for example when assuming that the photon leakage occurs only during the injection of the state into the cavity. The decoherence can then be modeled as a series of beamsplitters \cite{montenegro2015entanglement}. We will not consider these modifications here, but instead solve the Lindblad master equation numerically and compute the Fisher information $I_F(t)$ for the resulting mixed state. 

In order to calculate $I_F(t)$ under decoherence, we separate Eq. (\ref{eq:CFI}) above into a dimensionful and dimensionless part by writing 
\begin{align} \label{eq:SimpleFI}
I_F( t) &=\left(  \frac{\partial \bar{g}}{\partial g}  \right)^2 \int \mathrm{d}x_\lambda \frac{1}{p(x_\lambda, g)} \left(\frac{\partial p(x_\lambda, g)}{\partial \bar{g}} \right)^2.
\end{align}
Here, $\partial\bar{g}/ \partial g = \cos{\theta} \sqrt{m/(2 \hbar \omega_m^3)}$ is a dimensionful prefactor. The remaining integral
\begin{equation} \label{eq:NumCFI}
\bar{I}_F(t) = \int \mathrm{d}x_\lambda \frac{1}{p(x_\lambda, g)} \left( \frac{\partial p(x_\lambda,g)}{\partial \bar{g}} \right)^2 ,
\end{equation} 
is now dimensionless and is what we will evaluate numerically. A final estimate for $\Delta g$ can then be obtained by multiplying the value for $\bar{I}_F$ by $\cos^2{\theta} m/(2\hbar \omega_m^3)$, but as this is only a rescaling we will present only the results for $\bar{I}_F$ for clarity. 

In all subsequent numerical evaluations, we will set $\bar{k} = \bar{g} = 1$ and $\alpha = 1$ (note the choice of $\alpha \in \mathbb{R}$). Larger values will cause the system to quickly grow numerically unstable due to the inclusion of non-linear terms such as $(a^\dagger a)^2$ in the evolution in Eq. (\ref{eq:Time}). While  $\bar{k} = 1$ is experimentally achievable with the right choice of parameters, we can justify setting  $\bar{g} = 1$ by noting that it physically corresponds to a heavily inclined cavity with $\theta \approx \pi/2$. With these values, the prefactor in Eq. (\ref{eq:SimpleFI}) becomes $1/g^2$, which means that the overall Fisher information will be small. Thus these numerical investigations should only be seen as a indication as to how decoherence will affect $I_F(t)$, and not as predictions for the sensitivity of a realised device. We shall later extrapolate from this to make a prediction for realistic systems. 

To evolve the system, we use the Python library \textit{Qutip} \cite{qutip} and a 4th order Runge-Kutta-Fehlberg method \cite{ fehlberg1969low}. See the supplemental material for details. The results can be found in Fig. \ref{fig:CFIMomentum} for a momentum measurements with $\lambda = \pi/2$ and $\alpha = 1$. A measurement with $\lambda = 0$ and $\alpha = i$ would show the same results. As can be seen, larger $\kappa$ do affect the CFI adversely, but $\kappa = 0.1$ still leaves about 10\% of the maximum value. Note that we have plotted the dimensionless $\bar{I}_F(t)$ and not $I_F(t)$ as we are only interested in the general behaviour of the Fisher information.

\textit{Measuring the leaking photons}. 
In practise, a homodyne measurement is performed by monitoring and measuring the photons that continuously leak from the cavity. Aside from the experimental considerations, such a scheme also negates part of the photon dissipation considered above. We briefly estimate the CFI obtained through such as setup by using a simplified model where a pure vacuum state of the environment $\ket{0}_E$ is added to our original state $\ket{\Psi(t)}_{CO}$, giving us the combined state $\ket{\Psi(t)}_{CO}\otimes \ket{0}_E$. We then add a rotating wave interaction term $H_I$ to the Hamiltonian $H_g$ in Eq. (\ref{eq:GravHamiltonian}), of the form
\begin{equation}
H_I = \gamma \left( a^\dagger c + a c^\dagger \right),
\end{equation}
where $\gamma$  is an interaction strength and $c$ and $c^\dagger$ are the creation and annihilation operators of the environment. The effect of this interaction Hamiltonian is to couple the cavity state to the environment which causes information about $g$ to slowly leak out from the cavity into $\ket{0}_E$. 

As before, we evolve the full state for a single photon $|\alpha|^2 = 1$ and with parameters $ \bar{k} = \bar{g} = 1$.  To maximise the CFI, we choose $\alpha \in \mathbb{R}$ and $\lambda = \pi/2$. The results can be found in Fig. \ref{fig:Leaky} for a rescaled coupling strength $\bar{\gamma} = 0.1$, where $\bar{\gamma} = \gamma/\omega_m$.   As evident from the plot, we suffer a factor $10^{-2}$ reduction in the information that can be extracted from the system.  Note also that the behaviour of $I_F(t)$ for this scenario will most likely also resemble a delta function centered around $t = 2\pi$ for realistic parameters.

\textit{Physical realization and discussion}. In this section we shall first calculate the ideal Fisher information for the three optomechanical systems considered above, and then discuss the experimental challenges and advantages to an optomechanical gravimeter.  As we here calculate the fundamental sensitivity, which is unlikely to be realised, we will only concern ourselves with order-of-magnitude estimates.

Starting with the Fabry-Perot cavity system, we choose a fully vertical cavity with $\theta = 0$ and use the following state-of-the-art experimental parameters: We choose a mass $m = 10^{-6}$  kg, oscillator frequency $\omega_m = 10^3$ Hz, cavity frequency $\omega_C = 10^{14}$ Hz, cavity length $L= 10^{-5}$ m and a photon number of $|\alpha|^2 = 10^6$. For these values, the rescaled coupling constant in Eq. (\ref{eq:MirrorCavity}) becomes $\bar{k}_{\mathrm{FP}} \approx 2.30$, which gives us a Fisher information of $I_F = 1.58 \times 10^{28}$ m$^{-2}$ s$^4$.  This implies a sensitivity of $\Delta g  \approx 7.96 \times 10^{-15}$ ms$^{-2}$. 

Next, we look at a levitated micro-object confined in an ion trap interacting with an optical cavity, as demonstrated very recently in Refs. \cite{Millen2015iontrap, fonseca2016nonlinear}. Again setting $\theta = 0$ for maximal effect, we use mass $m = 10^{-14}$ kg, oscillator frequency $\omega_m = 10^2$ Hz, cavity frequency $\omega_C = 10^{14}$ Hz, volume $V = 10^{-18}$ m$^3$, cavity mode volume $V_C = 10^{-14}$ m$^3$, electric permitivity $\epsilon = 5.7$ for nanodiamonds, laser wavelength $\lambda = 1064\times 10^{-9}$ m and a photon number of $|\alpha|^2 = 10^6$. From these values we obtain $\bar{k}_{\mathrm{ND}} = 1963$  which leads to $I_F \approx  1.15 \times 10^{29}$ m$^{-2}$s$^4$. This gives us a final sensitivity of $\Delta g  \approx 2.94 \times 10^{-15}$ ms$^{-2}$ for levitated nanospheres.

Finally, let us also consider the cold atoms in a cavity. Based on \cite{brennecke2008cavity}, we choose the following parameters: A wavelength $\lambda = 780$ nm, implying $\omega_C = 10^{15}$ Hz, a single-atom coupling of $g_0 = 10^7 $ Hz, an atomic oscillation frequency $\omega_m = 10^2$ Hz, a single-atom mass $m = 10^{-25}$ kg, a detuning of $\Delta_{ca} = 10^{11}$ Hz, and a laser wavevector of $k_l = 10^8$ m$^{-1}$.  With $N =10^5$ atoms trapped in the cavity, we find that $\bar{k}_{BEC} = 2.30 \times 10^6$ and $I_F \approx 1.58 \times 10^{19}$ m$^{-2}$s$^4$, giving a sensitivity of $\Delta g \approx 2.5 \times 10^{-10}$ ms$^{-2}$. The reason for this disparity seems to be that the polarisability of the collection of cold atoms is not high enough to match the polarisability exhibited by the nanosphere. The number of trapped atoms can hardly match the number of atoms in a single nanosphere. One would either have to increase the number of atoms trapped in the cavity or increase the single-atom coupling strength to increase the Fisher information.

Let us briefly compare the results obtained here with the performance of other quantum systems in the literature. In Table \ref{tab:ComparisonExp} we have listed a variety of experimentally implemented gravimeters with their best achieved sensitivity to date. Table \ref{tab:ComparisonTheory}, on the other hand, lists the ideal fundamental limits to sensitivities calculated in this work and others. The values for $\Delta g$ and $\Delta g/ \sqrt{\mathrm{Hz}}$ are presented in units of ms$^{-2}$ and ms$^{-2}$ Hz$^{-1/2}$ respectively. The last column in Table \ref{tab:ComparisonExp} lists the integration time for each experiment, whereas in Table \ref{tab:ComparisonTheory} the last column lists the experimental cycle time, set by the oscillation frequency of the system in question. For atom interferometry, it is suggested in  \cite{chiow2011102} that sensitivities of $\Delta g \sim 10^{-12}$ ms$^{-2}$ might be achieved, and a study of the fundamental limits has very recently been presented in \cite{kritsotakis2017optimal}.

\begin{table}[h]
\caption{ A comparison of the sensitivity $\Delta g$ in  ms$^{-2}$ and $\sqrt{\mbox{Hz}}$-noise achieved by various gravimeter experiments, including the integration time.}
\begin{ruledtabular}
\begin{tabular}{Sl Sc Sc Sc}
\multicolumn{4}{Sc}{\textbf{Experiments }}\\  
 \textbf{System} & $\Delta g$  & $\Delta g/\mathrm{\sqrt{Hz}} $&  \textbf{Int. time}\\ 
\hline 
  LaCoste FG5-X \cite{LaCoste2016}  & $  1\times 10^{-9}$ & $1.5\times 10^{-7}$ &6.25 hours \\
  Atom intf.  \cite{hu2013demonstration} & $5 \times 10^{-9}$ & $4.2\times 10^{-8}$ & $100$ s \\
On-chip BEC \cite{abend2016atom}  & $7.8\times 10^{-10}$  & $ 5.3\times10^{-9}$ & $100$ s  \\
Optomech. accel. \cite{Cervantes2014} & $3.10\times 10^{-5}$  & $9.81 \times 10^{-7}$   & $10^{-3}$ s
\end{tabular} \label{tab:ComparisonExp}
\end{ruledtabular}
\end{table}

\begin{table}[h]
\caption{ A comparison of theoretical predictions for the ideal sensitivity $\Delta g$ in ms$^{-2}$ and $\sqrt{\mbox{Hz}}$-noise for various quantum systems. Values calculated in this work are denoted by *.  }
\begin{ruledtabular}
\begin{tabular}{Sl Sc Sc Sc}
\multicolumn{4}{Sc}{\textbf{Theoretical predictions}}
\\ 
 \textbf{System} & $\Delta g$  & $\Delta g/\mathrm{\sqrt{Hz}} $&  \textbf{Cycle time}\\  \hline
Magnetomech. \cite{johnsson2016macroscopic} & $2.2 \times 10^{-7}$ & $2.2 \times 10^{-9}$ & $10^{-4}$ s \\
Fabry-Perot optomech.* & $10 ^{-15}$ & $10^{-16}$ & $10^{-3}$ s\\
Levitated optomech.*  & $10^{-15}$ & $10^{-16}$ & $10^{-2}$  s\\	
Cold atoms* & $10^{-10}$ & $10^{-11}$ & $10^{-2}$ s\\
\end{tabular} \label{tab:ComparisonTheory}
\end{ruledtabular}
\end{table}
Let us now address some of the experimental challenges related to this scheme. Due to measurement inefficiencies and additional sources of decoherence not considered here, the final performance of optomechanical systems is of course expected to be lower than the values presented in Table \ref{tab:ComparisonTheory}.  While we have shown that the initial optomechanical state does not need cooling to the ground-state, thermal noise due to external influences \emph{during} the evolution will gradually decohere the oscillator motion. We estimate that in the case of a Fabry-Perot cavity cooled to a temperature of mK, a number of $\hbar \omega_m/(k_B T_{\mathrm{th}}) = N$ phonons are present in the system at any time, where $k_B$ is Boltzmann's constant and $T_{\mathrm{th}}$ is the system's temperature. To retain coherence throughout the evolution, we require that $\kappa_m N \ll \omega_m $, where $\kappa_m$ is the phonon dissipation rate. In other words, the timescale of phonon decoherence $\kappa_m$ must be much less than the characteristic timescale of the system. With $\omega_m = 1$ kHz, as we assumed for Fabry-Perot cavities, we find $N = 10^5$ and $\kappa_m = 10^{-2}$ Hz. A cavity which achieves such a decoherence rate must have a mechanical $Q$-factor of at least $Q = \omega_m/\kappa_m \sim 10^6$ to retain coherence, a regime which is not unprecedented. 

Let us now discuss what other parameters we require in order to realise this scheme experimentally. Most importantly, as the Fisher information ultimately scales with the oscillator frequency $\omega_m^{-5}$ we require $\omega_m$ to be small. At the same time we require the photon dissipation rate $\kappa$ to be as low as possible. This combination is difficult to achieve as low $\omega_m$ means the cavity must remain coherent over longer timescales. 
Therefore, the main experimental challenge of this scheme is to reduce $\omega_m$ and $\kappa$ at the same time. Taking our numerical results as guidance, we essentially require that $\bar{\kappa} = \kappa/\omega_m \ll 1$, which is nothing but the resolved sideband regime \cite{chan2011laser}. One of the best coherence times to date with a cavity linewidth of $\kappa = 660$ was demonstrated in \cite{zhao2009vibration}. Let us investigate the magnitude of the resulting $I_F$ for such a system: To achieve $\bar{\kappa} = 0.1$ given $\kappa = 660$, we set $\omega_m = 6600$ Hz and use $L = 9.4$ cm as reported in the paper. We keep $m = 10^{-6}$ kg and let $\omega_C = 10^{14}$ Hz as before. Because the oscillation frequency is quite high, we choose to calculate $I_F$ for the Fabry-Perot cavity with a mechanical mirror, as this performed slightly better for higher $\omega_m$. The resulting coupling constant is rather small with $\bar{k} = 1.44\times 10^{-5}$, and the Fisher information becomes $I_F \approx 2.16 \times 10^{15}$ m$^{-2}$s$^4$. This leads to $\Delta g \approx 2.15 \times 10^{-8}$ ms$^{-2}$. If we now assume that decoherence causes a similar proportion of the Fisher information to dissipate at these parameters compared to the ones chosen in our numerical simulations, we see that we retain about 10\% of the pure-state Fisher information. Using this assumption, we find $\Delta g \approx 6.80\times 10^{-8}$ ms$^{-2}$ and a $\sqrt{\mathrm{Hz}}$-noise of $8.37\times 10^{-10}$ ms$^{-2}$/$\sqrt{\mathrm{Hz}}$. This is directly comparable with the values in Table \ref{tab:ComparisonExp}, and so we believe that this scheme can be experimentally realised, although the challenges are not insignificant. 

Let us briefly discuss ways in which we can decrease $\kappa$ further and how this might affect the Fisher information. A heuristic estimate for $\kappa$ can be given by considering the number of times per second that a  single photon traverses the cavity. Each time the photon is reflected at the mirror, it has a $T = 1 - R$ chance of being transmitted instead of reflected. The photon bounces off a mirror $c/L$ times per second. Thus we can take the dissipation rate to be $\kappa = T c/L$, which means that increasing $L$ decrease the photon dissipation rate $\kappa$, as the photon is effectively spending longer inside the cavity. However, increasing $L$ also decreases the single-photon coupling constant, as we saw in the calculation above. This is true for all couplings we quote here, but it is perhaps most clearly seen for the case of the mechanical mirror and a Fabry-Perot cavity, with $k_{\mathrm{FP}}$ given by Eq. (\ref{eq:MirrorCavity}). $k_{\mathrm{FP}} $ scales with $L^{-1}$, and so do the other couplings, through their dependence on the cavity volume $V_C$ or the single-photon coupling $g_0$. We recall that the Fisher information depends on $\bar{k}^2$, which means that it ultimately scales with $L^{-2}$. Thus, changing $L$ by an order of $10$ will decrease the Fisher information by an order of $10^2$. This contributes to the challenges of realising this scheme. However, it is important to note that there are realistic ways of increasing $L$ without changing the single-photon coupling: One such method was explored in \cite{pontin2018levitated}, where $L$ was increased by adding an optical fibre to the cavity. 

Furthermore, in the above we proved the optimality of a homodyne detection scheme, but we also found that such a measurement must be performed within a rather narrow temporal window, of timescale $1/k$. Let us here estimate how quickly these measurements have to be performed based on the values we calculated for the coupling constant $k$. The nanospheres displayed the highest single-photon coupling $\bar{k}_{\mathrm{ND}} \times \omega_m$.  For the choice of $\omega_m = 10^2$ Hz, we find that measurements must be performed within $10^{-5}$ s, so we require at most microsecond precision. In comparison, we calculated $\bar{k}_{\mathrm{FP}} = 2.30$, which allows for a very comfortable $\approx 0.19$ s window. Both these measurement speeds are perfectly achievable.

In spite of these challenges, optomechanical systems come with a number of advantages. They can remain stationary while performing the measurement, in contrast to on-chip BECs or BEC fountains which need to be launched, and the short cycle time allows for a large number of measurements to be performed very quickly. An additional point which we did not elaborate on above is that the spatial resolution of optomechanical systems will be extremely high since the oscillator is displaced only by a minuscule distance. As a result, it will be possible to determine very fine local variations in $g$, something which is not possible using larger systems. The scheme presented in this work also allows for the creation of macroscopic spatial superpositions, which, as pointed out in \cite{johnsson2016macroscopic}, is of great interest to testing gravitational collapse models (see for example \cite{diosi1989models, hu2014gravitational, pfister2016universal}). 

Before we conclude, let us now briefly discuss the underlying physical differences between atom interferometry and the optomechanical interferometry described for the purpose of gravimetry. We estimate that the QFI for atom interferometry is given by $H_Q(T) \sim   n^2 T^4 k_C^2$  (see the supplemental material) up to an unknown geometric factor, where $n$ is the number of photons that deliver a momentum kick to the atoms, $T$ is the total time over which the gravimetric phase is accumulated, and $k_C$ is the laser wavevector. If we compare this to the Fisher information for optomechanical systems, we find that the Fabry-Perot cavity has a QFI that is larger by an enhancement factor $\xi_{\mathrm{FP}} \sim c^2/(n L^2\omega_m^2)$. This is due to the cavity confinement, whereby each photon interacts with the oscillator $c/(2L\omega_m)$ times per oscillation cycle, which is also the time period over which the gravimetric phase is accumulated. For the levitated nano-sphere, we find a $\xi_{\mathrm{Lev}} =\xi_{\mathrm{FP}} P^2 /(\epsilon_0 V_C  )^2$, where, again, for a micro-object containing $\sim 10^{13}$ atoms, the polarizability $P$ is much higher than that of a single atom. In practice, however, both of the enhancement factors will be damped by a factor $\sim 1/(\omega_m T)^4$ with respect to atom interferometry as the time of atomic interferometry $T$ is typically larger than the time $1/\omega_m$ of our scheme. Thus the sensitivity $\Delta g$ is seen to improve by a factor of $ \sqrt{n} L \omega_m^3 T^2/c   \sim \sqrt{n} \times 10^{-4}$ in our optomechanical scheme with respect to atomic interferometers. As $n$ increases, the differences level out. Strictly speaking, the enhancement is valid for when the cavity field remains coherent for the time $1/\omega_m$ over which our phase accumulation, i.e., $\kappa \ll \omega_m$ (the resolved side-band regime). However, our numerical results indicate that even in the presence of finite decoherence, say, $\kappa \sim 0.1 \omega_m$, the Fisher information is lowered only by a factor of about $10$ compared to the case of loss-less cavities. Finally, we can also compare the treatment presented in this work to a position measurement of a classical oscillator that has been displaced due to gravity. While a classical treatment of the problem returns a preliminary measurement sensitivity similar to what we have derived in this work, it fails to take into account effects such as radiation pressure and the full quantum nature of the cavity field. Most importantly, a classical treatment does not utilise the coherent nature of the oscillator, which as we saw above negates any initial thermal noise in the state.

\textit{Conclusion}.
In this work, we investigated a new scheme for measurements of the gravitational acceleration $g$ using a compact cavity optomechanical system with the usual trilinear optomechanical coupling to the cavity field. We derived a fundamental limit to the sensitivity $\Delta g$ by computing the quantum Fisher information and showed that the optimal sensitivity is achieved by a homodyne detection scheme performed on the cavity state at $t = 2\pi$. That is, no direct measurement of the mechanical oscillator is required. Using the expression in Eq. (\ref{eq:CFI}) and state-of-the-art experimental parameters, we predict a upper bound on the sensitivity of order $\Delta g \sim 10^{-15}$ ms$^{-2}$ for both a Fabry-Perot cavity and a levitated nanosphere cavity. This value compares favourably to all other currently available experimental and theoretical gravimetry proposals (see Table \ref{tab:ComparisonExp} and \ref{tab:ComparisonTheory}). Furthermore, the quantum nature of the oscillator ensures that any thermal distribution in its initial state does not affect the fundamental sensitivity. However, as our scheme relies on superpositions involving distinct coherent states, we require thermal decoherence \emph{during} one period of the oscillator motion to be negligible, which we estimate requires a $Q$-factor  of at least $10^6$ for the case of a Fabry-Perot cavity. To explore the effects of photons leaking from the cavity, we numerically explored a narrow parameter range with $\bar{k} = \bar{g} = 1$, which physically corresponds to a nearly horizontally aligned cavity. We found that this form of decoherence does affect the system's performance, but not severely. Finally, we briefly investigated what proportion of $\Delta g$ we retain by performing measurements on the photons that leak from the cavity. Using a simplified noise model, we found a reduction of $10^{-2}$ in the resulting Fisher information. 

After completing this work, the authors became aware of similar work carried out by Armata, Latmiral, Plato and Kim \cite{armata2017quantum}. 

The authors would like to thank Abolfazl Bayat, Nathana\"{e}l Bullier, Victor Montenegro, Dennis Schlippert, Stephen Stopyra and Doug Plato for useful comments and discussions. This work was supported by the EPSRC Centre for Doctoral Training in Delivering Quantum Technologies and the EPSRC grant EP/N031105/1.

\begin{appendix}

\section{Supplemental material} \label{sec:Appendix}
In this supplemental material, we provide additional analysis of concepts presented in the main text of the paper. We compute the linear entropy for the optomechanical state to better understand its behaviour, and perform some additional numerical calculations of the classical Fisher information within a broader parameter range. We then elaborate on the numerical methods used to calculate the Fisher information for mixed states and comment on their numerical stability. Furthermore, we present a short section on the simplified noise model used to estimate the Fisher information for measurements on leaking photons. Finally we derive an estimate for the quantum Fisher information for atom interferometry gravimetry and we also suggest a scheme for how an optomechanical system can surpass the standard Heisenberg limit. The code used for most of the numerical computations in this paper can be found at \url{https://github.com/sqvarfort/Coherent-states-Fisher-information}

\subsection{Fisher information for thermal states}
In this section, we will show that the Fisher information remains unchanged if the initial state is a thermal coherent state. Experimentally, this corresponds to a system that has not been cooled to the ground state. 

A general thermal mixed state is given by 
\begin{equation}
\rho(\xi, \beta) = \frac{1}{Z} \sum_{m = 0}^\infty e^{- \hbar \beta \omega} D(\xi) \ket{m}\bra{m} D^\dagger(\xi), 
\end{equation}
where $D(\xi) = e^{\xi b^\dagger - \xi^* b}$ are Weyl displacement operators and $\beta$ is now the inverse temperature rather than the coherent state parameter. The partition function is given by $Z = \frac{1}{1 - e^{- \hbar \beta \omega}} $. We must now show that the initial state
\begin{equation}
\rho = \ket{\alpha}\bra{\alpha}  \otimes \frac{1}{Z} \sum_{m = 0}^\infty e^{- \hbar \beta \omega} D(\xi) \ket{m}\bra{m} D^\dagger(\xi), 
\end{equation}
still decouples under the dynamics. Applying the time-evolution operator in Eq. \ref{eq:Time} gives the state
\begin{align}
\rho(t) &= \sum_{n,n'} \frac{\alpha^n (\alpha^*)^{n'}}{\sqrt{n! n'!}} e^{i [\bar{k}^2 (n^2 - n^{'2}) - 2 \bar{k}\bar{g}(n-n')] \tau} \ket{n}\bra{n'}  \nonumber \\
&\times \frac{1}{Z} \sum_{m} e^{-  \hbar \beta \omega} e^{[ (\bar{k} (n-n') -\bar{g}) \eta \xi^* e^{it} - (\bar{k} (n-n') - \bar{g}) \eta^* \xi e^{-it}]/2} \nonumber  \\
&\times e^{( \varphi_n(t) b^\dagger - \varphi^*_n(t)  b} \ket{m}\bra{m}  e^{\varphi^*_n(t)  b - \varphi_n(t)  b^\dagger }, 
\end{align}
with $\varphi_n(t) = (\bar{k} n - \bar{g} ) \eta + \xi e^{-it} $. Now, at $t = 2\pi$, we know that $(\bar{k} n - \bar{g} ) \eta + \xi e^{-it} = \xi$, and $\eta = 1- e^{-it}  = 0$, which means that the above state simplifies to 
\begin{align}
\rho(t) &= \sum_{n,n'} \frac{\alpha^n (\alpha^*)^{n'}}{\sqrt{n! n'!}} e^{i [\bar{k}^2 (n^2 - n^{'2}) - 2 \bar{k}\bar{g}(n-n')] 2\pi}  \ket{n}\bra{n'}  \nonumber \\
&\otimes\frac{1}{Z} \sum_{m} e^{-  \hbar \beta \omega}  \nonumber  e^{ \xi b^\dagger - \xi^* b} \ket{m}\bra{m}  e^{\xi^*  b - \xi  b^\dagger }, 
\end{align}
where we can see that the oscillator has returned to its original state and has become completely decoupled. The cavity state has the same form as in Eq. \ref{eq:state} and so the resulting Fisher information will be the same as the one we calculated for coherent states in Eq. (\ref{eq:QuantumFisher}).

\subsection{Linear entropy}
\begin{figure}[h]
\subfloat[ \label{fig:EntropyA}]{%
  \includegraphics[width=0.49\linewidth]{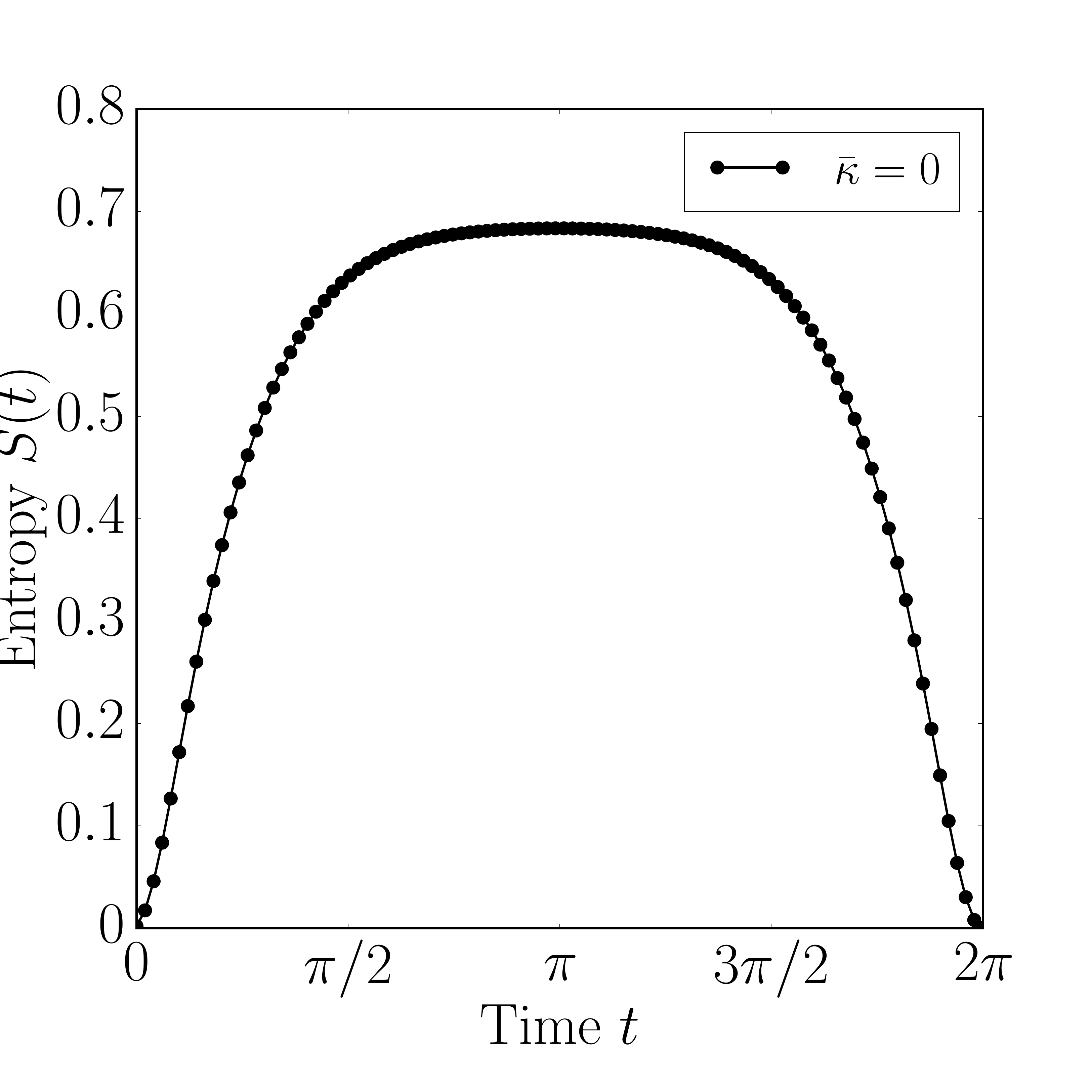}%
}\hfill
\subfloat[ \label{fig:EntropyB}]{%
  \includegraphics[width=0.49\linewidth]{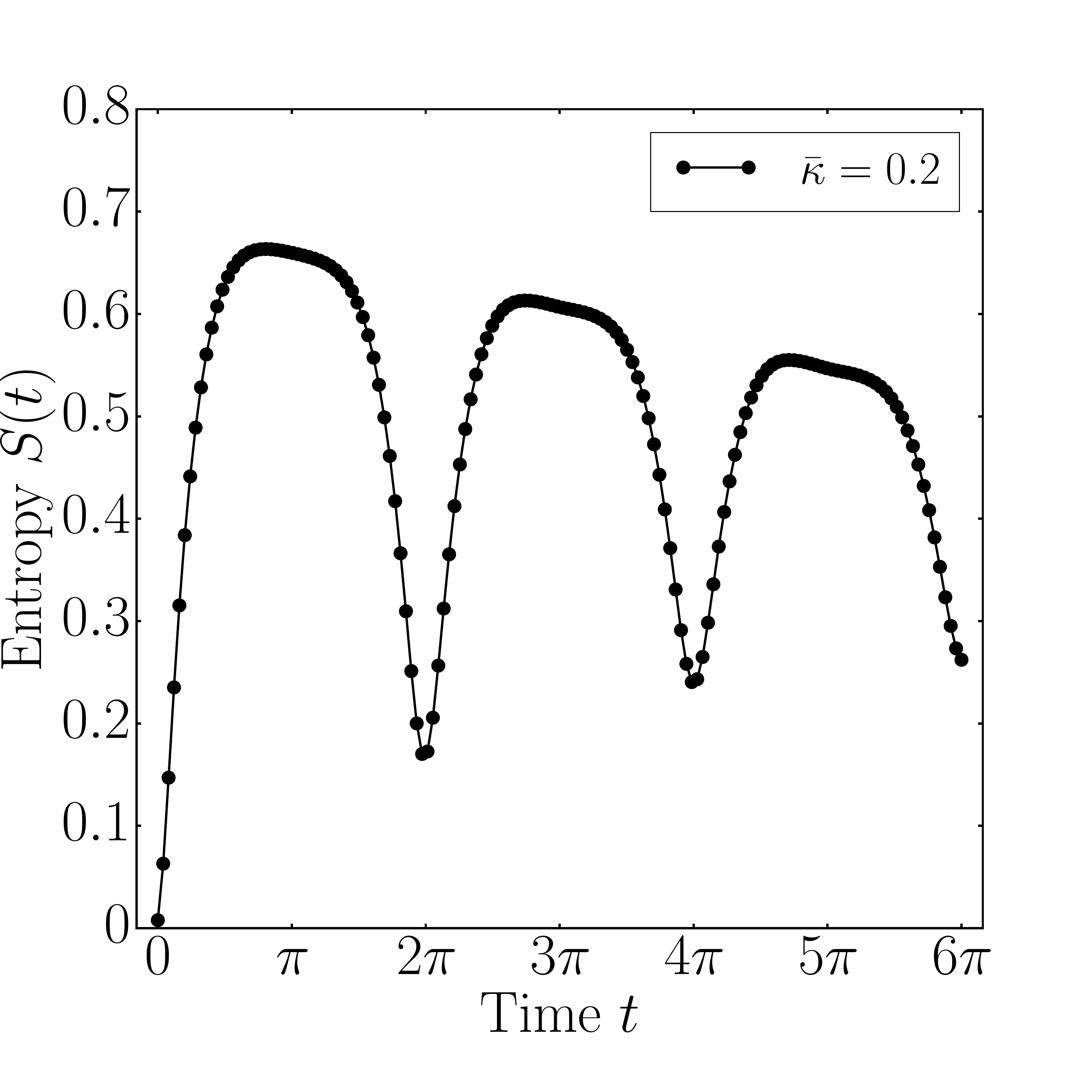}%
} \hfill 
\caption{Plots showing the linear entropy $S(t)$ for (a) a pure cavity state and (b) a decohering cavity state with $\bar{\kappa} = 0.2$. }
\label{fig:Entropy}
\end{figure}

We noted that the light and mechanics periodically entangle and disentangle under the gravitational Hamiltonian. In order to see this more clearly, we can compute the linear entropy $S(t)$ for the traced out cavity state in Eq. (\ref{eq:TracedCavity}). The linear entropy is defined as 
\begin{equation}
S(t) =  1 - \trace{[\rho^2_L(t)]}. 
\end{equation}
The linear entropy tells us about the entanglement between the cavity and oscillator states. The results can be found  in Fig. \ref{fig:EntropyA} for pure state and in Fig. \ref{fig:EntropyB} for states undergoing decoherence due to photons leaking from the cavity. We see that $S(t) $ increases until the state is maximally entangled at $t = \pi$. While a pure state completely decouples the light and mechanics at $t = 2\pi$, a decohering state becomes increasingly mixed. Note that the decoupling for pure states at $t = 2\pi $ occurs regardless of the values chosen for $\bar{k}$ and $\bar{g}$. 

\subsection{Additional notes on the CFI for pure states}
In the main text we computed the classical Fisher information (CFI) for pure states using the expression in Eq. (\ref{eq:CFIHomodyne}). 

\begin{figure}[h]
\subfloat[ \label{fig:FisherPosition}]{%
  \includegraphics[width=0.49\linewidth]{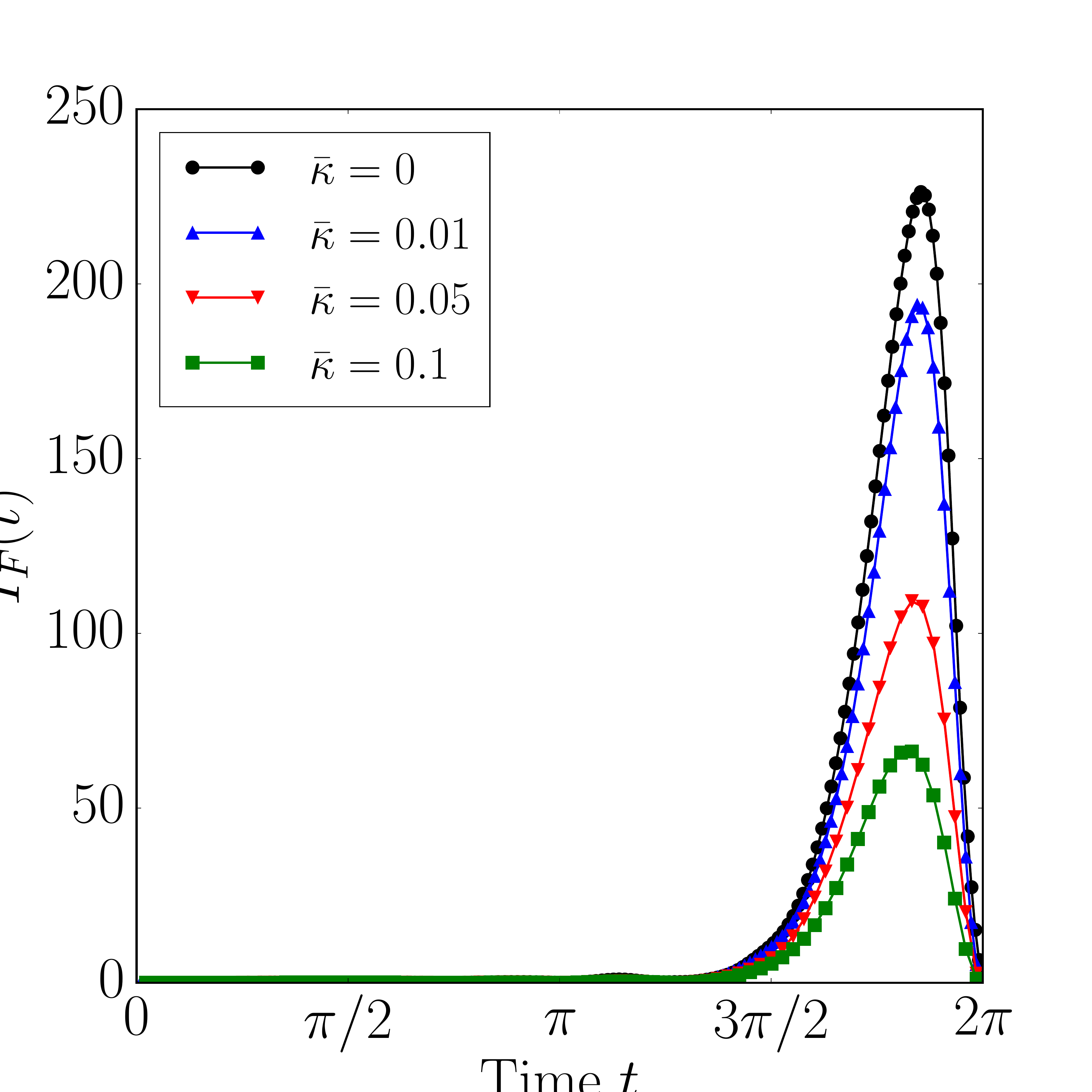}%
}\hfill
\subfloat[ \label{fig:CFIb}]{%
  \includegraphics[width=0.49\linewidth]{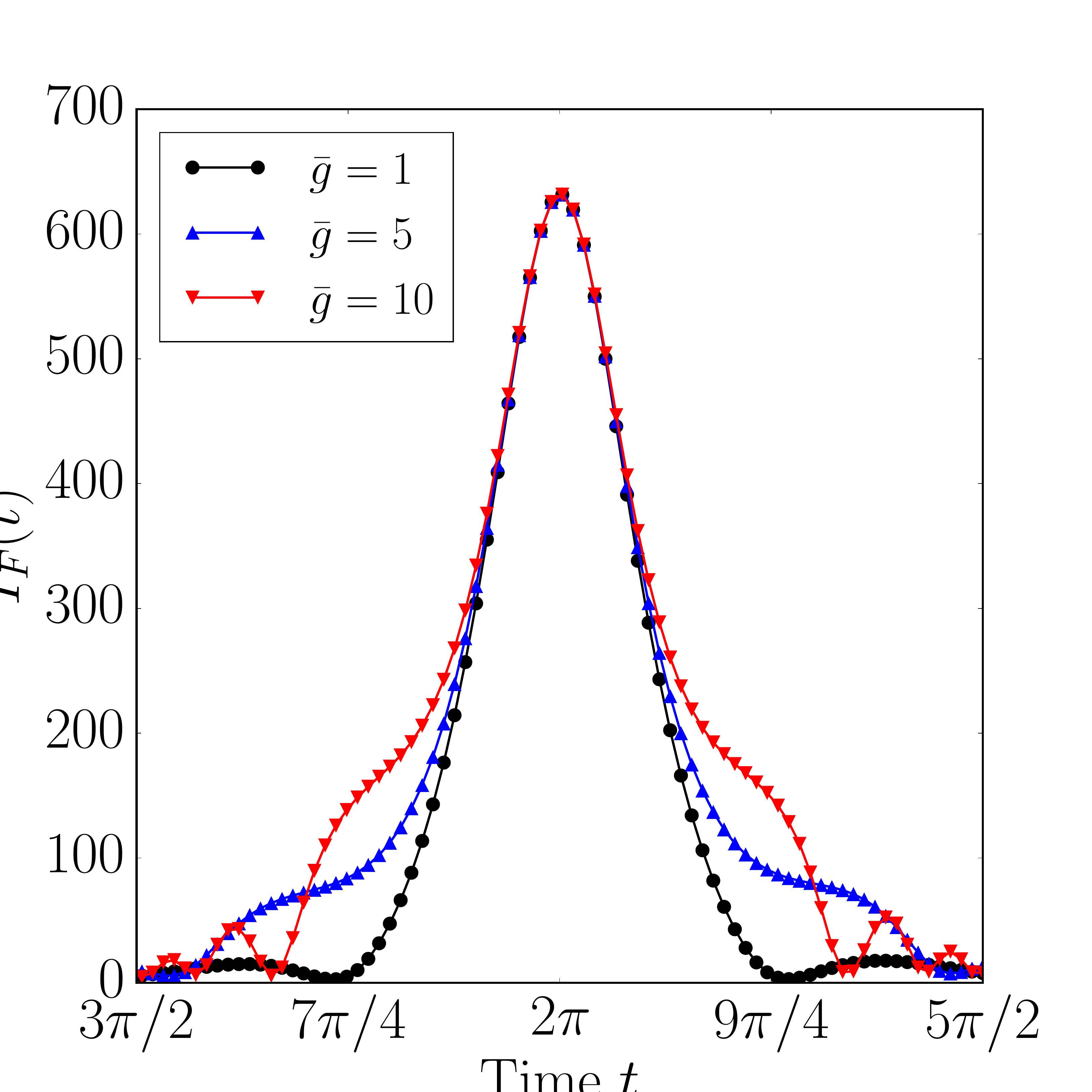}%
} \hfill 
\caption{Plots showing the Fisher information for a homodyne measurement with (a) $\lambda = 0$, $\bar{k} = \bar{g} = 1$ with various decoherence rates, and (b) the Fisher information for $\lambda = \pi/2$ with $\bar{k} = 1$ and $\bar{g} = 1,5,10$. }
\label{fig:VaryParameters}
\end{figure}

Let us now investigate the effects of the parameters $\bar{k}$ and $\bar{g}$ on the CFI. In the main text, we plotted the values of $I_F$ at $t = 2\pi$. For small enough $\bar{k}$, the Fisher information is not entirely suppressed for values $t\neq 2\pi$. If for $\bar{k} = \bar{g} = \alpha = 1$ we measure the position quadrature (with $\lambda = 0$) instead of momentum, we find that the Fisher information peaks just before $t = 2\pi$. This can be seen in Figure \ref{fig:FisherPosition}. However, comparing this value  with the QFI at the same time ($t = 5.82$ shows that it is much lower at $I_F = 2.37$ ms$^{-2}$ versus $QFI = 628$ ms$^{-2}$. The numerical value in the plot is $\bar{I}_F = 228$, but recall that we have to add the dimensionful prefactor of $\cos^2{\theta} m/(2\hbar \omega_m^2)$, which for $\bar{g} = 1$ becomes just $1/g^2$. Thus we see clearly that the homodyne measurement is optimal only at $t = 2\pi$.

We showed in Eq. (\ref{eq:HomodyneFisher}) that the CFI is independent of $\bar{g}$  and scales with $\bar{k}^2$ at $t = 2\pi$.  To emphasize this point, we here provide some additional computations of the CFI for different values of $\bar{k} $ and $\bar{g}$. Note that we are again computing the dimensionless part of $I_F$ in Eq. (\ref{eq:CFIHomodyne}), leaving out the prefactor $\cos^2{\theta}  m /(2 \hbar \omega_m^3)$ for clarity. 

Fig. \ref{fig:CFIa} shows the CFI for $\bar{k} = 1,2,3$ with $\bar{g} = 1$, and Fig. \ref{fig:CFIb} shows its behaviour for $\bar{g} = 1,2,3$ with $\bar{k} = 1$. As expected, the CFI scales with $\bar{k}^2$ at $t = 2\pi$. We also note some additional oscillations near the peak as $\bar{k}$ becomes larger.  For larger values of $\bar{g}$, we see only marginal changes in the behaviour of the function in the region around $t = 2\pi$. 

\subsection{Numerical methods and stability}
If we wish to compute the CFI for states undergoing decoherence, Eq. (\ref{eq:CFIHomodyne}) is no longer valid and we must evolve the state numerically. We do so by calculating the dimensionless part $\bar{I}_F$ in Eq. (\ref{eq:NumCFI}) for a mixed state $\rho(t)$. The probability distribution $p(x, g)$ is easy to obtain numerically, since we can solve for the eigenstates of any matrix operator and use these as our POVM elements. That is, we can easily define a position operator $\hat{x}_C$ as a finie-dimensional matrix and solve for its eigenstates.  The non-trivial part of this computation concerns obtaining the derivative $\partial p(x, g)/\partial \bar{g}$. 

Calculating the derivative numerically can be done in a number of ways. The simplest one is to use a higher order method of the central difference theorem. We obtained good and accurate results with the 4th-order five-point method. It is given by 
\begin{align} \label{eq:NumDerivative}
f'(x) = \frac{1}{12 h} \biggl( &- f(x+ 2h) + 8 f(x+h)  \nonumber \\
&- 8f(x-h) + f(x- 2h) \biggr) + \mathcal{O}(h^4). 
\end{align}
The disadvantage of using the five-point method is that it requires five runs of the computation, each with a slightly altered parameter. In our case, it means that we have to evolve the state five times, each with slight changes to the parameter $g$. For pure states, this is fine, as each simulation is completed within minutes.  If we wish to add decoherence, however, we must work with the density matrix. As the number of matrix elements grows with $N^4$, where $N$ is the dimension of a single Hilbert space, this soon becomes extremely computationally expensive. 

In the end however, this was our preferred numerical method as computing the CFI can still be done within reasonable time-scales using the optimised master equation solver provided by the \textit{Qutip} library \cite{qutip}. Note however that this method yields two different sources of numerical errors: Errors that originate in the numerical method used to solve the Lindblad equation and errors that originate from the cut-off in the numerical derivative. 

In order to double-check the obtained results, we made use of another method which provides an exam estimate for the differentiation. Let us here present the method in detail. We start by noting that we can write the derivative of the probability distribution as
\begin{equation}
\frac{\partial p(g, x)}{\partial\bar{g}} = \trace{ \left[ \frac{\partial \rho(g)}{\partial \bar{g}} \Pi_x \right]}, 
\end{equation}
which holds provided the POVM elements $\Pi_x$ do not depend on $g$. Note that we are differentiating with respect to $\bar{g}$ instead of $g$ and that we have suppressed the dependence of $t$ for clarity. This statement also holds for subsystems of $\rho(g)$, which we can see by noting that the derivative distributes over a joint separable system $\rho_{AB}= \rho_A \otimes \rho_B$ as
\begin{equation}
\frac{\partial \rho_{AB}}{\partial \bar{g}} = \frac{\partial \rho_A}{\partial \bar{g}} \otimes \rho_B + \rho_A \otimes \frac{\partial \rho_B}{\partial  \bar{g}}. 
\end{equation} 
Performing a measurement with $\Pi_x$ that only acts on subsystem $A$ then gives
\begin{align}
\mathrm{tr}_B \left[ \frac{\partial \rho_{AB} }{\partial \bar{g}} \Pi_x \right] &=  \mathrm{tr}_B\left[ \frac{ \partial \rho_A}{\partial \bar{g}} \Pi_x \otimes \rho_B \right]  \nonumber \\
&\quad+   \mathrm{tr}_B\left[ \rho_A \Pi_x \otimes \frac{\partial \rho_B}{\partial \bar{g}}  \right] . 
\end{align}
The second term reduces to zero because $\trace{[\partial_{\bar{g}} \rho_B]} = \partial_{\bar{g}} \trace{[\rho_B]} = 0$. While we have shown this for separable states, the same argument can be extended to entangled states by linearity. 

In order to obtain the evolution for this state, we must now solve a modified version of the master equation.  That is, given the Lindblad equation in Eq. (\ref{eq:Lindblad}), 
\begin{equation}
\dot{\rho}(\bar{g}) = - \frac{i}{\hbar} \left[H(\bar{g}), \rho(\bar{g}) \right] + L \rho(\bar{g}) L^\dagger - \frac{1}{2} \{\rho(\bar{g}),  L^\dagger L \}, 
\end{equation}
where $\{\cdot , \cdot \}$ denotes the anti-commutator, we now differentiate with respect to $\bar{g}$ to obtain
\begin{align} \label{eq:ModMasterEq}
\partial_{\bar{g}} \dot{\rho}(\bar{g}) = &- \frac{i}{\hbar} \left[ \partial_{\bar{g}} H(\bar{g}), \rho(\bar{g}) \right] - \frac{i}{\hbar} \left[ H(\bar{g}) , \partial_{\bar{g}} \rho(\bar{g}) \right]   \nonumber \\
&- \frac{1}{2} \{ \partial_{\bar{g}} \rho(\bar{g}) , L^\dagger L\}, 
\end{align}
where we have used the notation $\partial_{\bar{g}} = \partial / \partial \bar{g}$. A more complicated form is obtained if the Lindblad operators $L$ depend on $\bar{g}$, which here is not the case. Taking the trace with some POVM element of this object  allows us to compute the derivatives of the probabilities using just one single run of the numerics. Ultimately, we are solving the following system of coupled differential equations:
\begin{align}
&\frac{d}{dt} \left(
\begin{matrix}
\rho  \\ \partial_{\bar{g}} \rho
\end{matrix} \right) \nonumber \\
&= 
\left(
\begin{matrix} - \frac{i}{\hbar} \left[ H, \rho \right]  +  \left( L \rho L^\dagger - \frac{1}{2} \{ \rho, L^\dagger L \} \right) \\
  - \frac{i}{\hbar} \left( \left[\partial_g H , \rho \right]  +\left[ H , \partial_{\bar{g}} \rho \right] \right)+  L \partial_{\bar{g}} \rho L^\dagger - \frac{1}{2} \{ \partial_{\bar{g}}, \rho L^\dagger L \} 
\end{matrix}
\right). 
\end{align}
A solution can be found using any standard higher-order method, such as the family of Runge-Kutta ODE solvers. Once the time-evolved state $\partial \rho/\partial \bar{g}$ has been obtained, we proceed as usual to compute the probability distribution and the CFI. With this method, we avoid round-off errors that appear in the five-point numerical derivative above.

Let us make a few remarks about what influences the stability of the simulation. We start by considering the nature of coherent states and how they are represented numerically. Coherent states have support on infinite Hilbert spaces, whereas numerically we must work with finite matrices. It is therefore necessary to introduce a cut-off in the dimension used to represent the state. This leads to a gradual loss of coherence as information is pushed beyond the cut-off. In other words, we use a finite Hilbert space $\mathcal{H}$, meaning that we truncate the space by letting $a^\dagger \ket{N-1} = 0$, where $\mathrm{dim}(\mathcal{H}) = N$. Furthermore, the appearance of $(a^\dagger a)^2$ in $U(t)$ causes the system to become anharmonic and numerical instabilities grow fast for Hilbert spaces with small dimension $N<50$. 

The information loss due to smaller Hilbert spaces is difficult to assess, since any good ODE solver will preserve the purity of the state throughout the simulation. Rather, it can be noted as a gradual deterioration of the trajectory in phase space, with the effect that states fail to return to their original position in phase space at $t = 2\pi$. That is, we require that $\langle \hat{x}(0) \rangle \approx \langle \hat{x}(2\pi) \rangle $ and $\langle \hat{p} (0) \rangle \approx \langle \hat{p}(2\pi) \rangle$ for the simulation to be deemed stable. 

The system dynamics depend strongly on the dimensionless constants $\bar{k}$ and $\bar{g}$. Larger $\bar{k}$ and $\bar{g}$ will cause the system to evolve more rapidly, as evident from their appearance in the phase of the state in Eq. (\ref{eq:state}). This in turn causes the numerical inaccuracies to accumulate more rapidly. When computing the CFI for mixed states, we restrict our investigations to the parameter range $\bar{k} = \bar{g} = 1$ for precisely this reason. 

Finally, it should be noted that we have not provided a full error estimate for any of the results computed here. Since we are only interested in the general behaviour of the CFI, small variations in the numerical estimates will not matter. 

\subsection{Measuring the leaking photons}
In the main text we presented a simplified noise model to estimate the Fisher information obtained from performing measurements on the leaking photons. The model is limited in its application because it involves a fully unitary process between the system and the environment.  In other words, at some later time $t$, all information about $g$ will be transferred back from the vacuum state into the coherent cavity state, a clearly unphysical process. Therefore, we limit ourselves to small values of $\gamma$, which also ensures the stability of the simulations. 

\begin{figure}[h]
\subfloat[ \label{fig:CFIb}]{%
  \includegraphics[width=0.49\linewidth]{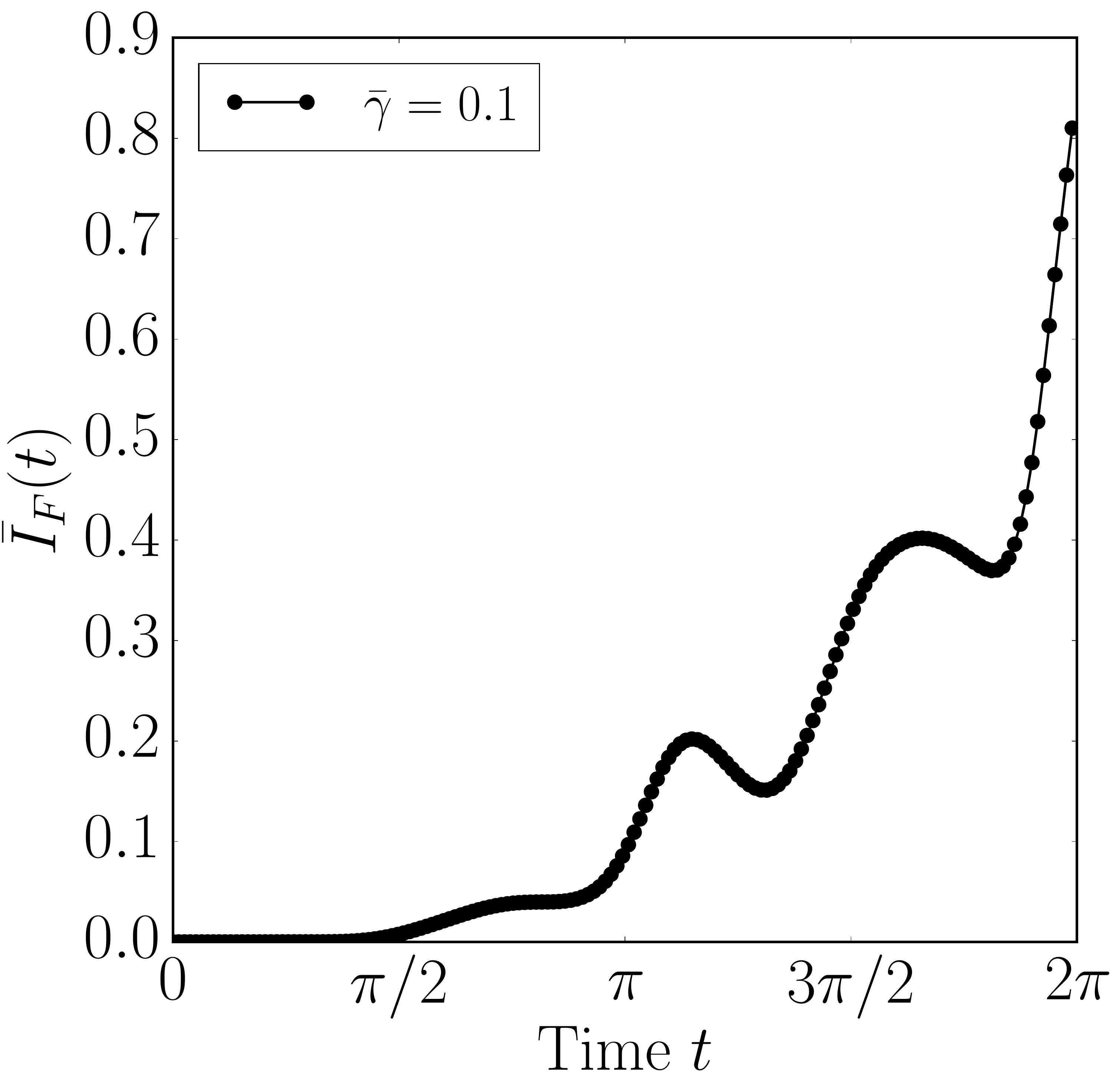}%
} \hfill 
\subfloat[ \label{fig:CFIb}]{%
  \includegraphics[width=0.49\linewidth]{Fisher_leaky01_momentum.pdf}%
} \hfill 
\subfloat[ \label{fig:CFIa}]{%
  \includegraphics[width=0.49\linewidth]{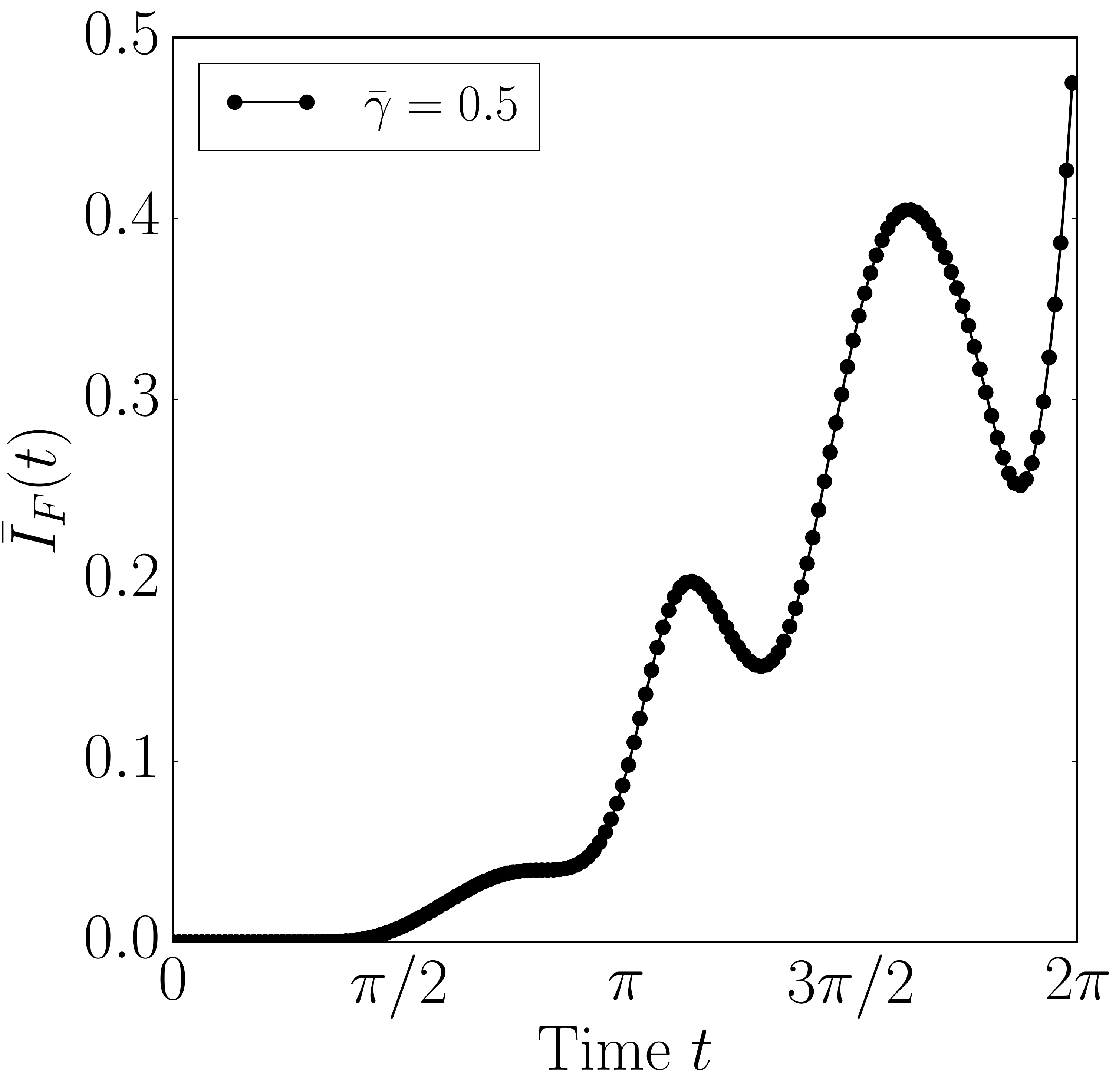}%
}\hfill
\subfloat[ \label{fig:CFIb}]{%
  \includegraphics[width=0.49\linewidth]{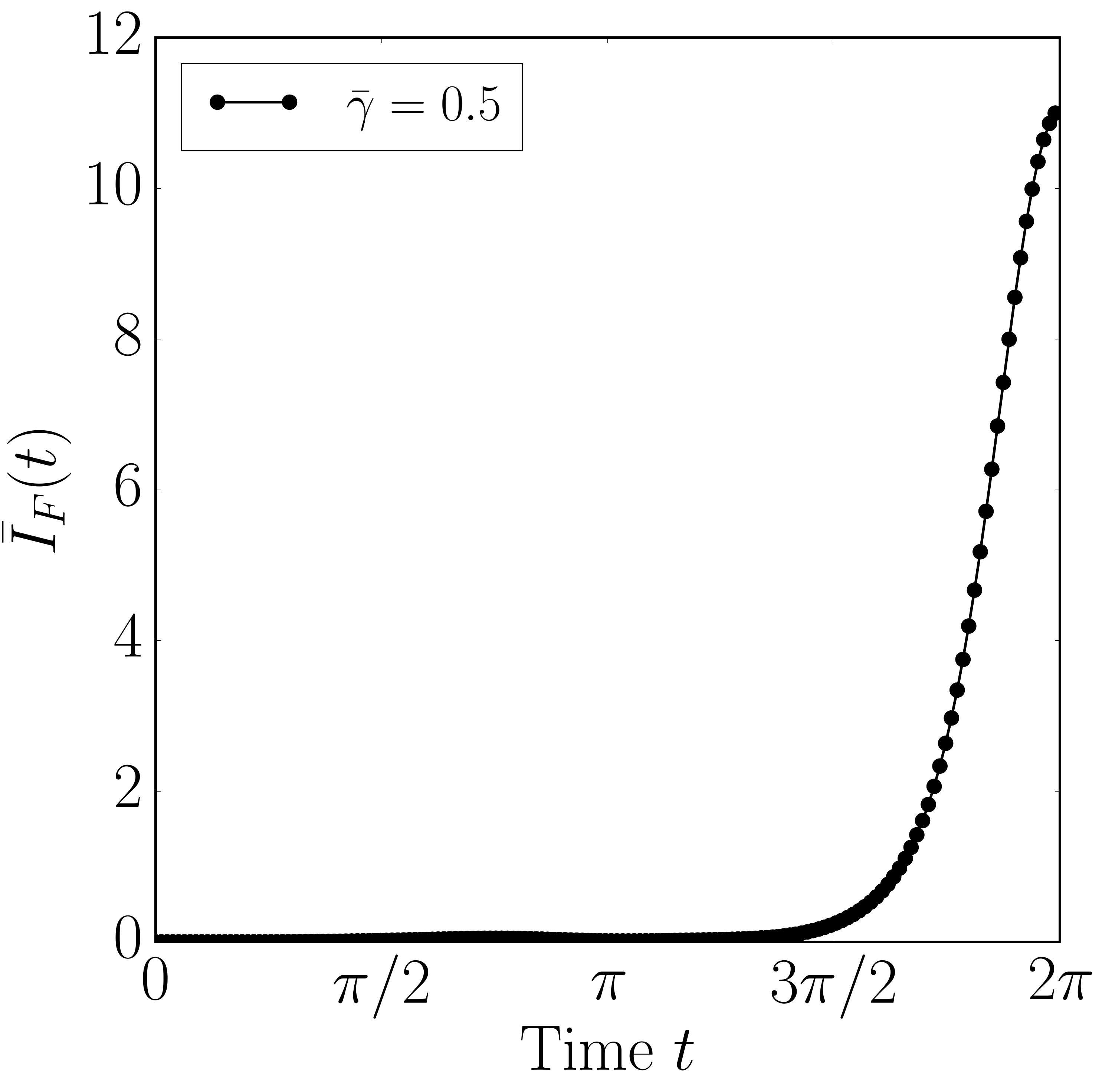}%
} \hfill 
\caption{Plots showing the classical Fisher information for measurements on the leaking photons. (a) and (b) show the CFI for position and momentum measurements respectively with $\gamma = 0.1$, and (c) and (d) show the CFI for position and momentum measurements with $\gamma = 0.5$.  The additional parameters were set to $\bar{k} = \bar{g} = 1$, $\beta = 1$ and $\alpha = 1$. }
\label{fig:LeakingPhotons}
\end{figure}

In Fig. \ref{fig:LeakingPhotons}, we present additional results from numerical computations with $\gamma = 0.1$ and $\gamma = 0.5$, including both the position and momentum quadrature measurements for each run. As before we used parameters $\bar{k} = \bar{g} = 1$ and $\alpha = 1$.  We note that a larger $\gamma$ does not significantly impact the extracted CFI, which is to be expected as the information about $g$ is generated by the motion of the oscillator and is transferred to the light field at a set rate not influenced by $\gamma$. With the choice of $\alpha \in \mathbb{R}$, we obtain largest $I_F$ for momentum measurements, although  at $t = 2\pi$ we do not see a complete reduction to zero for position measurements. Clearly the addition of the environment changes the behaviour of the CFI. The difference in $I_F$ for the position measurements between the two runs is possibly due to numerical errors.

\subsection{Fisher information for atom interferometry}
In the main text, we stated that the Fisher information for atom interferometry is equal to $n^2 T^4 k^2_C$ and that optomechanical systems are enhanced by a factor of $\xi_{\mathrm{FP}} \sim  c^2 /(n L^2 \omega_m^2)$ in comparison. Here we detail the derivation of the expression and the enhancement factor. 

In atom interferometry, we prepare the atoms in a superposition of a ground state $\ket{g}$ and an excited state $\ket{e}$, such that the full state becomes $\ket{\psi} = (\ket{g} + \ket{e} )/\sqrt{2}$. Photons are then used to separate the two states by momentum transfer, causing them to take two different paths through a gravitational potential. We then assign a potential gravitational energy $ mg \Delta x$ for the excited state, where $\Delta x$ is the difference in height between the two paths. The phase accumulated by the excited state is then equal to $e^{i mg \Delta  T/\hbar}$, where $m$ is the atomic mass and $T$ is the time of flight. We must now determine $\Delta x$. Ignoring any geometric factors associated with the paths, we assume that the distance roughly depends on the atoms' velocity $v$ and their time of flight $T$. That is, we let  $\Delta x \sim v T$. The total velocity is determined by the momentum transfer from the photons in the laser pulse, and is therefore proportional to the number of photons $n$. The momentum carried by one photon is given by $\hbar k_C$, where $k_C$ is the wavevector of the photon (which we take to be the same as the wavevector of the photons in the cavity). Thus, assuming that each photon transfers all of its momentum to the atom, we find that 
\begin{equation}
\Delta x \sim vT \sim \frac{n\hbar k_C }{m} T. 
\end{equation}
If we insert this into the expression for the phase and apply it to the state, we find 
\begin{equation}
\ket{\psi} = \frac{1}{\sqrt{2}} \left( \ket{g} + e^{ i ngk_C T^2} \ket{e} \right).
\end{equation}
Calculating the quantum Fisher information for this state is straight-forward. We find that
\begin{align}
H_Q &= 4 \left( \langle \partial_g \psi | \partial_g \psi \rangle - |\langle \partial_g \psi | \psi \rangle |^2 \right) \nonumber \\
&= 4 \left( \frac{n^2 k_C^2 T^4}{2} - \frac{n^2 k_C^2 T^4}{4} \right) \nonumber \\
&= n^2 k_C^2 T^4. 
\end{align}
Since $k_C$ has dimension m$^{-1}$, this expression has the correct units of s$^4$m$^{-2}$. In terms of scalability, we note that this expression surpasses the Heisenberg limit in terms of the number of photons $n$, and that it is highly dependent on the time of flight $T$.

We can now compare this with the optomechanical Fisher information for the Fabry-Perot cavity. The explicit Fisher information with $k_{\mathrm{FP}}$ inserted into Eq. (\ref{eq:QuantumFisher}) given by 
\begin{equation}
H_{Q, \mathrm{FP}} = \frac{32\pi^2n  \cos^2{\theta}}{\omega_0^6} \frac{\omega_C^2}{L^2}. 
\end{equation}
where we have replaced $|\alpha|^2$ by $n$. To compare the two expressions, we let $\omega_m^4 \sim 1/T^4$,  $\omega_C =2\pi c /\lambda  $, and $k_C = 2\pi/\lambda$. We set $\theta = 0$ for clarity and then divide them to find that 
\begin{align}
\xi_{\mathrm{FP}} &=  \frac{ 32 \pi^2 n c^2T^4/(\omega_m^2\lambda^2 L^2) }{ 4\pi^2n^2  T^4 / \lambda^2 } \nonumber \sim \frac{ c^2}{n \omega_m^2L^2}. 
\end{align}
This is the enhancement factor mentioned before. A similar analysis can be performed for the levitated nanospheres.

\subsection{The Heisenberg limit}
The QFI obtained in Eq. (\ref{eq:QuantumFisher}) scales linearly with the number of photons $|\alpha|^2$, and so does not allow us to reach the so-called Heisenberg limit. To achieve a quantum speedup in the sensitivity $\Delta g$, one must instead show that the Fisher information scales with the number of photons squared, that is $|\alpha|^4$. As we have shown, a coherent state will not achieve this, but a highly non-classical superposition of two Fock states will be more successful. We will show this by considering the following state
\begin{equation}
\ket{\varphi} = \frac{1}{\sqrt{2}} \left( \ket{0} + \ket{n} \right) \ket{\beta}, 
\end{equation}
where $n$ denotes the number of photons. This state evolves under $H_g$ into
\begin{align}
\ket{\varphi(t)} = \frac{1}{\sqrt{2}} ( \ket{0} + e^{2\pi i (\bar{k}^2 n^2 - 2 \bar{k} \bar{g}n )} \ket{n} )\ket{\beta}.
\end{align}
Finally the QFI obtained at $t = 2\pi$ for this state is given by
\begin{align}
H_Q(2\pi) &= 4\left( \langle \partial_g \varphi | \partial_g \varphi \rangle - |\langle \partial_g \varphi | \varphi \rangle|^2 \right) \nonumber \\
&= 4\left( \frac{\partial \bar{g}}{\partial g} \right)^2 \left( 8\pi^2 \bar{k}^2 n^2  - 4 \pi^2 \bar{k}^2 n^2 \right) \nonumber \\
&= 16\cos^2{\theta} \frac{m}{ \hbar \omega_m^3} \pi^2 \bar{k}^2 n^2 , 
\end{align}
where the appearance of $n^2$ indicates that the standard Heisenberg limit has been surpassed. 

While these states are very difficult to prepare, using $\bar{k}_{\mathrm{FP}} = 2.30$ for the Fabry-Perot mirror and cavity systems, and choosing a single photon with $n = 1$ gives us a sensitivity $\Delta g \sim 1.5\times 10^{-11}$ ms$^{-2}$. Although it is not as high as the $\Delta g$ obtained for coherent states, the increased power of the scaling offers interesting opportunities as technologies improve. 

\end{appendix}

\bibliography{bibliography}

\end{document}